\documentclass[prl,superscriptaddress,twocolumn,notitlepage,longbibliography]{revtex4-1}
\usepackage[unicode=true, bookmarks=true,bookmarksnumbered=false,bookmarksopen=false,
 breaklinks=false,pdfborder={0 0 1},backref=false,colorlinks=true]
 {hyperref}
\hypersetup{
linkcolor=magenta,urlcolor=blue,citecolor=blue,pdfstartview={FitH},urlcolor=blue}

\usepackage{amsfonts}
\usepackage{subfigure}
\usepackage{amsmath}
\usepackage{amssymb}
\usepackage{amsbsy} 
\usepackage{epsfig}
\usepackage{graphicx}
\usepackage{epstopdf}
\usepackage{bbm}

\def\Im{\text{Im}}

\def\be{\begin{equation}} \def\ee{\end{equation}}
\def\bea{\begin{eqnarray}} \def\eea{\end{eqnarray}}

\begin{document}

\title{Scale-free localization and $PT$ symmetry breaking from local non-Hermiticity}

\author{Bo Li}
\affiliation{
Institute for Advanced Study, Tsinghua University, Beijing, 100084, China}

\author{He-Ran Wang}
\affiliation{
Institute for Advanced Study, Tsinghua University, Beijing, 100084, China}

\author{Fei Song}
\affiliation{
Institute for Advanced Study, Tsinghua University, Beijing, 100084, China}

\author{Zhong Wang} 
\affiliation{
Institute for Advanced Study, Tsinghua University, Beijing, 100084, China}
\date{\today}

\begin{abstract}
We show that a local non-Hermitian perturbation in a Hermitian lattice system generically induces scale-free localization for the continuous-spectrum eigenstates. When the perturbation lies at a finite distance to the boundary, the scale-free eigenstates are promoted to exponentially localized modes, whose number is proportional to the distance. Furthermore, when the local non-Hermitian perturbation respects parity-time ($PT$) symmetry, the $PT$ symmetry breaking is always accompanied by the emergence of scale-free or exponential localization. Intriguingly, we find a concise band-structure condition, which tells not only when the continuous-spectrum $PT$ breaking of scale-free modes can occur, but also the precise $PT$-breaking energy window. Our results uncover a series of unexpected generic phenomena induced by a local non-Hermitian perturbation, which has interesting interplay with $PT$ symmetry.

\end{abstract}

\maketitle
\textit{Introduction}. Non-Hermiticity of a Hamiltonian often arises in open or non-equilibrium systems~\cite{Ashida2020,Sternheim1972}. It induces remarkable phenomena such as unidirectional invisibility~\cite{lin2011unidirectional, Feng2013}, single-mode lasing \cite{hodaei2014PT, feng2014singlemode}, and enhanced sensitivity~\cite{Hodaei2017,Chen2016,Liu2016,Chen2018}. Many of them are related to the parity-time ($PT$) symmetric Hamiltonians, which generally have two phases when their parameters are varied, namely the $PT$-exact and $PT$-broken phases, for which the eigen-energies are real and complex, respectively ~\cite{bender1998real, Bender_2007,Ozdemir2019, El-Ganainy2018, Miri2019EP}. They are bridged by the $PT$ breaking transition at the exceptional points (EPs).

Recently, progress in non-Hermitian topological phases has revealed the non-Hermitian skin effect (NHSE)~\cite{yao2018edge,yao2018chern,kunst2018biorthogonal,lee2018anatomy,Longhi2019Probing,Helbig2019NHSE,xiao2020non,alvarez2017,Ghatak2019NHSE,Wang2022morphing,Bergholtz2021RMP}, the aggregation of eigenstates near boundaries, which underlines a striking modification of bulk-boundary correspondence in non-Hermitian systems. Whereas NHSE stems from the non-Hermiticity of the entire system, another common scenario in non-Hermitian physics is that the system is subject to local non-Hermiticity. For example, it is often the case that the gain or loss occurs only at the boundaries, preserving the Hermiticity of the bulk system~\cite{Landi2021,Federico2022,Wiersig2018,Heinrich2020, Alba2022,datta1997electronic,Burke2020localNH,Dora2021Friedeloscillations}. The system can be modeled by a Hermitian Hamiltonian with local non-Hermitian terms.

In this paper, we show that, in one-dimensional Hermitian lattice systems, local non-Hermiticity at boundaries generally gives rise to a scale-free localization, for which the spatial decay length of eigenstates is proportional to the system size, such that the eigenstates retain the same profiles if the system size is taken as the measure of length. The phenomenon is robust and generic. As such, it fundamentally differs from the scale-free NHSE under global non-Hermiticity, which requires fine tuning certain parameters ~\cite{Li2020,Li2021,Kazuki2021}. More surprisingly, when the local non-Hermiticity is located at a finite distance to boundary, it also brings about a collection of bound states that exponentially localize to the impurity. The localization length and population of bound modes rely on the distance but are independent of the system size. Furthermore, when the local non-Hermiticity respects $PT$ symmetry, we find that the emergence of scale-free or bound states always coincide with $PT$ breaking. Unexpectedly, the $PT$ breaking associated with scale-free modes under open boundary conditions (OBC) is dictated by a simple band-structure-based criterion, which also tells the energy window of $PT$ breaking.



\begin{figure}
\includegraphics[width=1\linewidth]{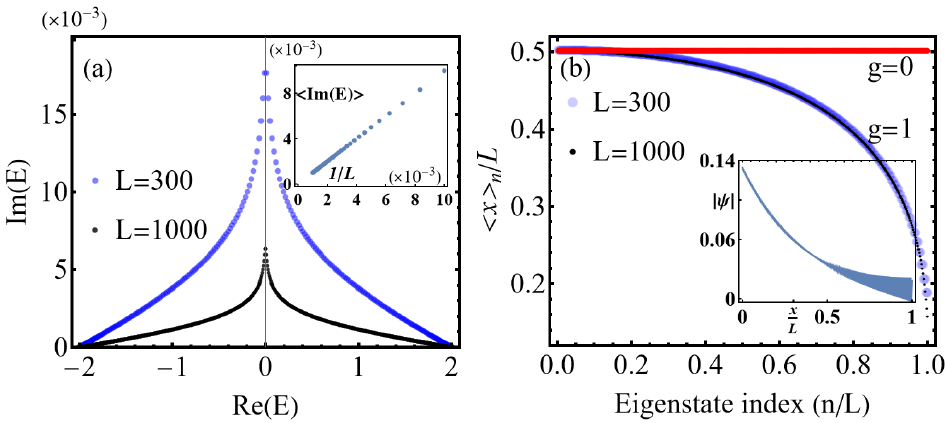}
 \caption{Scale-free localization for model Eq.~\eqref{eq:toymodel}. (a) Eigenvalues in the complex plane with $g=1$. The inset shows the averaged imaginary part of energy $\langle\Im(E)\rangle$ for varying $1/L$. (b) Eigenvalue-resolved mean positions. The eigenstate index is arranged in the ascending order of the eigenvalue imaginary part. The inset gives a typical wave function profile exhibiting scale-free localization. For (a) and (b), $t=1$.}\label{fig:scale_free}
\end{figure}

\textit{Scale-free localization from local non-Hermitian perturbation}. To be concrete, we start with a simple model:
\begin{eqnarray}\label{eq:toymodel}
H=\sum_{j=1}^{L-1} t(|j\rangle\langle j+1|+|j+1\rangle\langle j|)+i g |1\rangle\langle 1|,
\end{eqnarray}
where the first two terms describe nearest hopping with a real parameter $t$, the last term represents a boundary gain controlled by $g$. OBC has been built in the Hamiltonian.  Let an eigenstate be $|\psi\rangle=\sum_j\psi_j|j\rangle$,  then the eigen-equation in the bulk reads $t(\psi_{j-1}+\psi_{j+1})=E\psi_{j}$,  whose characteristic equation is given by $t(\beta+\beta^{-1})=E$. This allows the eigenstate ansatz:
\begin{eqnarray}\label{eq:ansatz}
\psi_j(\beta)=c_1\beta^j+c_2\beta^{-j},
\end{eqnarray}
where $c_1$, $c_2$ are coefficients determined by boundary conditions. Applying the ansatz to the real-space eigen-equation $H|\psi(\beta)\rangle=E|\psi(\beta)\rangle$, OBC results in a zero determinant condition:
\begin{eqnarray}\label{eq:boundarycondition}
\det\left[
\begin{array}{cc}
ig\beta-t & ig\beta^{-1}-t\\
t\beta^{L+1} &t\beta^{-(L+1)}
\end{array}
\right]=0.
\end{eqnarray}
The spectrum of the Hamiltonian Eq.~\eqref{eq:toymodel} is complex [Fig. \ref{fig:scale_free}(a)], which demands non-unitary solutions, i.e., $|\beta|\neq 1$, since $|\beta|=1$ always gives real eigenvalues. If $|\beta|^L\gg 1$ for a large system ($L\gg 1$), Eq.~\eqref{eq:boundarycondition} reduces to $t\beta^{L+1}(ig\beta^{-1}-t)=0$, which could be satisfied by a bound state solution $\beta\simeq ig/t$ when $|g|>t$ ; the case for $|\beta|^L\ll 1$ is similar. 
However, the Hamiltonian has $L$ eigenvalues; thus, the rest of the solutions should scale as $|\beta|^L\sim O(1)$ even for $L\gg 1$. Therefore, to the leading order of $L^{-1}$, the solution approximately satisfies
\begin{eqnarray}\label{beta}
|\beta|\simeq e^{c/L},
\end{eqnarray}
where $c$ is size independent (it is generally eigenstate-dependent), so that $|\beta|^L=e^c\sim O(1)$. This displays a scale-free localization, i.e., the decay length is proportional to the system size. The (right) eigenstate profile respects a scale invariance  $|\psi_L(x)|\simeq e^{c x/L}\simeq|\psi_{sL}(sx)|$, with $s$ being a scaling factor. This feature can be further demonstrated by a eigenvalue-resolved mean position $\langle x\rangle_n=\sum_{j=1}^L|\psi_{n,j}|^2 j/\sum_{j=1}^L|\psi_{n,j}|^2$. As plotted in  Fig.~\ref{fig:scale_free} (b), $\langle x\rangle_n$ for the non-Hermitian case ($g=1$) deviates from the uniform distribution ($g=0$),  and the curve shows a perfect self-similarity upon varying system size (with appropriately scaling the coordinates and eigenvalue index). In addition, it is evident that the imaginary part of eigenvalues are proportional inversely to the system size $L$ because $\text{Im} E \propto |\beta|-|\beta^{-1}|\approx 2 c/L$, which agrees with Fig.~\ref{fig:scale_free} (a). From our analysis, it is clear that the scale-free localization is a generic consequence of a local non-Hermitian perturbation to a Hermitian lattice Hamiltonian~\cite{Guo2023scalefree,distinction}.

\begin{figure}
\centering
\includegraphics[width=1\linewidth]{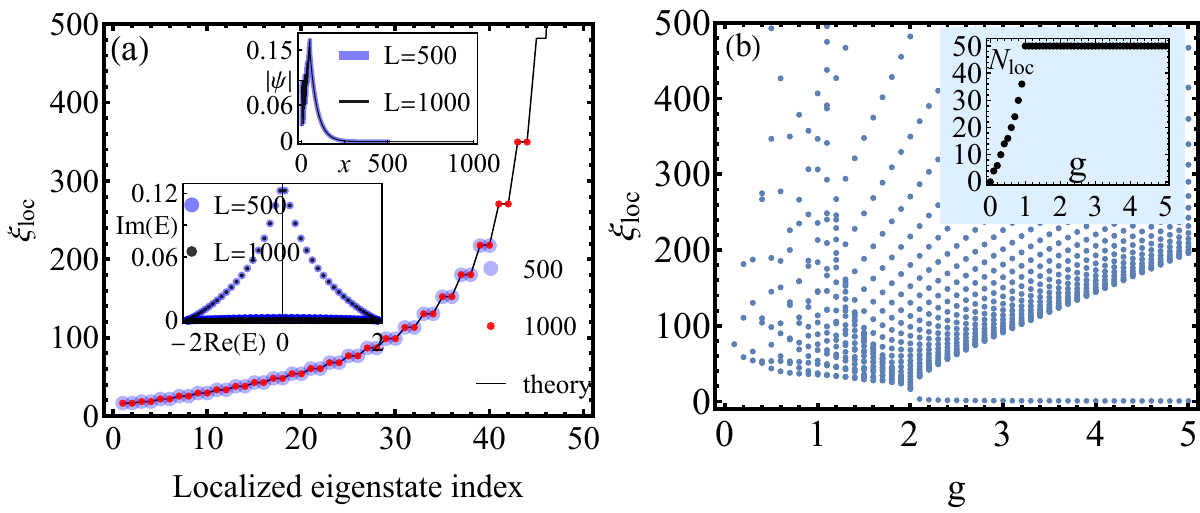}
\caption{(a) Localization length extracted from exactly diagonalized wave function and theory, where only $\xi_{\text{loc}}\leq 500$ is shown; the insets show exemplified eigenvalues and eigenstate profile for different system size. $t=1, d=50$, and $g=2$ are used. (b) Localization length ($\xi_{\text{loc}}\leq 500$) and the number of bound states (the inset) as a function of impurity strength $g$, $t=1, d=50$.}
\label{fig:localized_state}
\end{figure}

\textit{Accumulation of bound states}. Moving the position of the non-Hermitian impurity away from the boundary, we unveil another interesting consequence: Aside from scale-free modes, a local (even single-site) non-Hermitian impurity can induce arbitrary number of exponentially localized modes. The localization length ($\xi_{\text{loc}}$) and population of these modes depend on the distance between the impurity and the boundary, but not on the system size. For concreteness, we replace the non-Hermitian term in the model in Eq.~\eqref{eq:toymodel} by $ig|d\rangle\langle d|$, meaning that the distance between impurity and boundary is $d$. The insets in Fig.~\ref{fig:localized_state} (a) show that there are a bunch of eigenmodes aggregating to the impurity, and they have a size-independent imaginary part of the eigenvalues (isolated from the continuous spectrum) and localization length. Like the case of Eq.~\eqref{eq:toymodel},  the eigenvalues here follow $E=t(\beta+\beta^{-1})$, while the eigenstate wave function takes the general form: $|\psi\rangle=\sum_{1\leq j<d}\psi_j^{(1)}|j\rangle+\sum_{d\leq j\leq L}\psi_j^{(2)}|j\rangle$, with $\psi^{(\nu)}_j=a_\nu\beta^{j}+b_\nu\beta^{-j}$ ($\nu=1,2$). By adapting the bulk equation to the boundary and the impurity in the thermodynamic limit ($L\rightarrow\infty$, but $1\ll d\ll L$), we obtain the equation for localized modes:
\begin{eqnarray}\label{eq:localizedmodes}
ig\beta^{2d+1}-t\beta^2-ig\beta+t=0,
\end{eqnarray}
in which $|\beta|<1$ is imposed to ensure the states being piled up to the impurity. Numerically solving this equation (taking $d=50$), we obtain $d$ localized modes, whose localization lengths agree well with that ($\xi_{\text{loc}}=1/|\ln|\beta||$) extracted from eigenstate wave functions for $L=500$ and $1000$, as shown in Fig.~\ref{fig:localized_state}. Moreover, Fig.~\ref{fig:localized_state} (b) shows that, while varying the strength of impurity can substantially affect the localization length, the exponential localization is obtained for a wide range of $g$.

Notably, exponentially localized modes from Eq.~\eqref{eq:localizedmodes} can be viewed as scale-free modes with respect to the subsystem between the boundary and impurity, given that the localization length is comparable with the length of the subsystem: $|\beta|\sim e^{c/d}$. Meanwhile, they are exponentially localized modes for the whole system since $d\ll L$. 
Moreover, the localized modes arising from local non-Hermiticity here differ sharply from the Hermitian counterpart [by removing $i$ in Eq.~\eqref{eq:localizedmodes}], where only a few localized modes are obtained (not proportional to $d$). 

\textit{$PT$ breaking and localization}. We now show that, if a system with local non-Hermitian perturbation respects $PT$ symmetry, an intriguing interplay arises between $PT$ breaking and scale-free (or exponential) localization. Let us consider a $PT$-symmetric Hamiltonian $H=H_0+V$, where $V$ is a local non-Hermitian perturbation lying at or near the boundaries, and $H_0$ is a one-dimensional (1D) Hermitian chain with $L$ sites. Specifically, $H_0$ contains $n$th nearest-neighbor hopping with parameter $t_n (\in\mathbbm{C})$ and can be well adapted to periodic boundary conditions (PBC) or OBC.
To solve the eigenstates and eigenvalues, one takes the wave function ansatz $\psi_j=\sum_{\mu=1}^{2M}c_\mu \beta_\mu^j$ ($c_\mu$ takes different values for intervals partitioned by in-bulk impurities), where $\beta_\mu$ satisfies the characteristic equation $E=\sum_{n=1}^M t_n\beta^n+t_{n}^\ast \beta^{-n}$,
with the same energy $E$. Legitimate $E$ is determined by the boundary condition and/or eigen-equations near impurities.
In the $PT$-exact phase, we know \textit{a prior} that $E$ must be real, so that at least two of the corresponding $\beta_\mu$'s are on the unit circle, i.e., $\beta=e^{ik}$ with real momentum $k$,  and their contributions dominate the wave function in the bulk~\cite{Yokomizo2019}. The real-valuedness of $E$ dictates that $V$ cannot drag $\beta$ away from the unit circle and therefore the eigenstates remain extended (see Ref.\cite{supplemental} for more details). In the $PT$-broken phase, however, complex energies enforce $|\beta|\neq 1$, which corresponds to either a scale-free or bound state. 
Therefore, we conclude that the $PT$ symmetry breaking arising from local non-Hermiticity is always accompanied by the emergence of scale-free or (and) exponential localization.

\begin{figure}
      \begin{tabular}{cc} \includegraphics[width=1\linewidth]{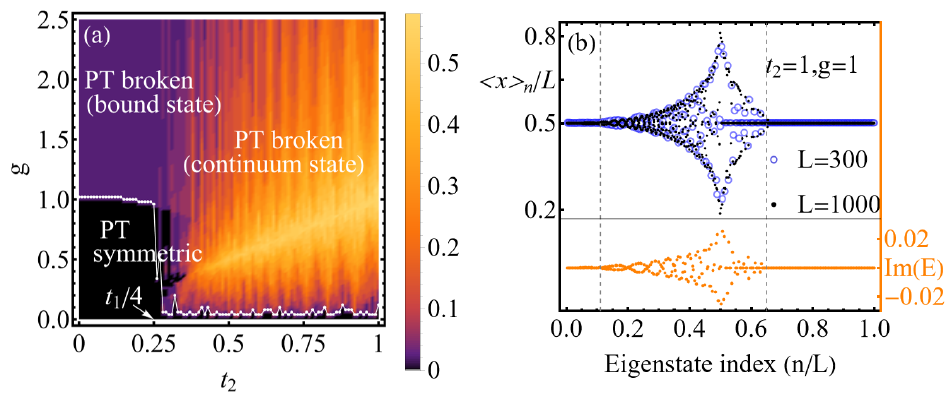}\\ \includegraphics[width=0.9\linewidth]{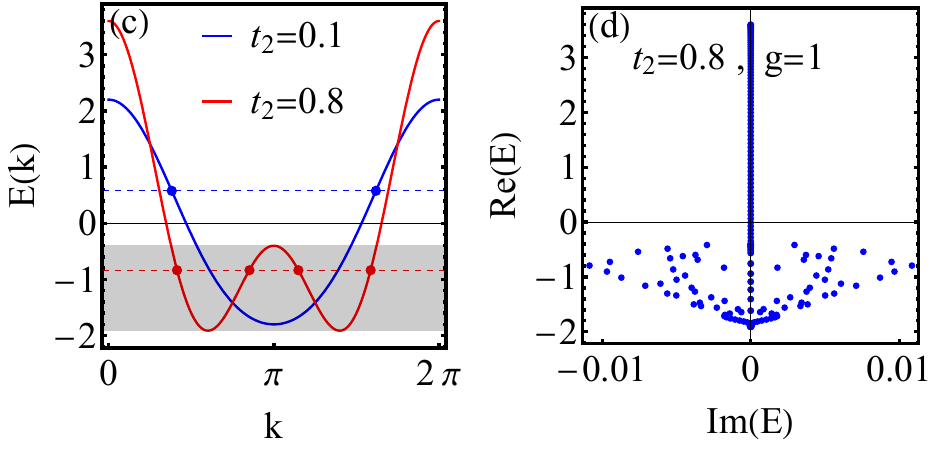}
      \end{tabular}
 \caption{ (a) Phase diagram for an open chain driven by boundary gain and loss. The color map shows $P_{\text{com}}$. The system size is $L=100$. (b) The eigenvalue-resolved mean position $\langle x\rangle_n$ (upper, left ticks, black) and  the imaginary part of eigenvalue (lower, right ticks, orange), where band index is arranged in the ascending order of eigenvalue real part. (c) Band structure of $H_0$. When $t_2>t_1/4$, the curve has two local minima and the shadow region covers the energy range that allows $PT$ breaking. (d) Eigenvaues in complex plane (with $L=100$). The real parts of complex eigenvalues fall into the shadowed energy range in (c). In all plots, $t_1=1$. }\label{fig:PTbreaking_OBC}
\end{figure}

\textit{$PT$ breaking criterion under OBC}. Furthermore, we find that $PT$ breaking associated with the scale-free modes, referred to as continuous-spectrum $PT$ breaking, has an unexpected interplay with boundary conditions. It turns out that the continuous-spectrum $PT$  breaking due to boundary non-Hermiticity occurs under PBC if the perturbation is strong enough~\cite{supplemental}. In sharp contrast, we find that, under OBC, the (unperturbed) band structure dictates the $PT$ breaking.

We state the conclusion first: For an open chain with band structure given by $\{E_n(k)\}$ (single or multiple band) in momentum space, the $PT$ symmetry breaking (of  continuous spectrum) arising from boundary non-Hermiticity can only take place at energies with more than one pair of equal-energy points. For instance, $PT$ breaking can occur for the system with the red band curve in Fig.~\ref{fig:PTbreaking_OBC} (c) in the shadowed energy window, where four equal-energy points exist, while it is prohibited for the blue band curve, as only two equal-energy points exist.

\begin{figure}
\includegraphics[width=1\linewidth]{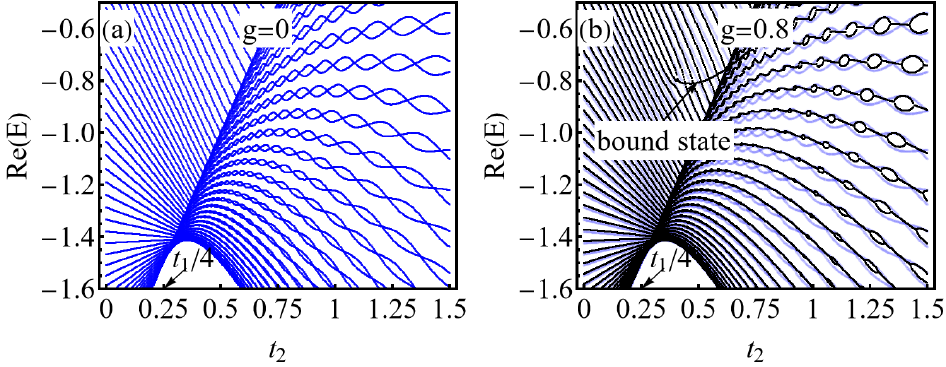}
 \caption{Spectrum for model described by Eqs.~\eqref{eq:OBCmodel1} and~\eqref{eq:OBCperturbation1}. (a) The Hermitian case. (b) The real part of  non-Hermitian spectrum (black) with $g=0.8$, overlapped with Hermitian spectrum (blue). The $PT$ breaking occurring between bound states is marked.  $t_1=1$ and $L=100$ are fixed.}\label{fig:spectrum}
\end{figure}

We demonstrate the mechanism in the following by assuming the single band, though extension to multi-band cases is straightforward. The physics can be understood by investigating the formation of EPs during $PT$ breaking, which requires the coalescence of at least two eigen-energies, see Fig.~\ref{fig:spectrum}. The eigenvalue is still real-valued at EP, so that it can be captured by the band structure $E(k)$, where the value of $k$ is determined by the local perturbation and boundary condition. Due to the energy degeneracy at EP, the band structure at $E(k)=\varepsilon_{\text{EP}}$ should be able to hold at least two eigenmodes, such that they develop a Jordan block at EP.  The eigenstates under OBC are standing waves composed of forward and backward plane wave, i.e., for a given energy $\varepsilon$, $\psi_{\text{OBC}}(x)\sim c_1 e^{ik_1 x}+c_2 e^{-ik_2 x}$ with $E(k_1)=E(-k_2)=\varepsilon$. Thus, only one OBC eigenmode can be constructed from two real solutions of the equation $E(k)=\varepsilon$, which manifests as the absence of eigenvalue degeneracy (e.g., see the spectrum for $t_2<t_1/4$ in Fig.~\ref{fig:spectrum}).  Therefore, EP can never exist at an energy with only one pair of equal-energy points, where only one eigenmode is allowed. In contrast, in the presence of more than one pair of equal-energy points, at least two eigenstates of the same energy exist, which enables EP formation.

As a comparison, the PBC case~\cite{supplemental} is different in the sense that each equal-energy point represents an independent plane-wave eigenmode $\psi_{\text{PBC}}(x)\sim e^{ikx}$, which allows EP formation from two equal-energy points. Note that our statement does not apply to isolated bound states, which are not captured by band structure $E(k)$ with real $k$.


According to our criterion, a single-band system with only nearest-neighbor hopping, whose band structure looks like the blue one in Fig.~\ref{fig:PTbreaking_OBC}(c), can not have continuous-spectrum $PT$ breaking under OBC, in stark contrast with the PBC behavior~\cite{supplemental}. This motivates us to consider a model with second-nearest-neighbor hopping
\begin{eqnarray}\label{eq:OBCmodel1}
	H_0=\sum_{j=1}^{L-1}t_1|j\rangle\langle j+1| + \sum_{j=1}^{L-2}t_2|j\rangle\langle j+2| +\text{H.c}.
\end{eqnarray}
where $t_1$, $t_2$ are real and positive, and the non-Hermitian boundary potential $V$ is given by
\begin{eqnarray}\label{eq:OBCperturbation1}
	V=i g ( |1\rangle\langle 1|- |L\rangle\langle L|).
\end{eqnarray}
The phase diagram is plotted in Fig.~\ref{fig:PTbreaking_OBC} (a), where the $PT$ breaking is quantified by the proportion of complex eigenvalues $P_{\text{com}}=n_{\text{com}}/L$, with $n_{\text{com}}$ being the number of complex eigenvalues. The $PT$ breaking of the continuous spectrum is prohibited until $t_2>t_1/4$ because the band structure has one local minimum for $t_2<t_1/4$, thus only allows one pair of equal-energy points. Figure~\ref{fig:PTbreaking_OBC} (c) gives two representative cases, where the red curve fulfills the requirement and the shadow region marks the allowed energy window for $PT$ breaking. As shown in (d), the energy real part of $PT$-broken modes precisely falls into this range. In addition, we confirmed the correspondence between $PT$ breaking and scale-free localization in Fig.~\ref{fig:PTbreaking_OBC} (b), where the eigenvalue-resolved mean position only shows deviation from middle position for $PT$-broken modes. It is also worth noting that there are two $PT$-broken modes (even for $t_2<t_1/4$) stemming from isolated bound states near boundaries if the non-Hermitian term is large enough, irrespective of the criterion for continuous spectrum.

The real part of the spectrum in Fig.~\ref{fig:spectrum} (d) shows that the $PT$ breaking takes place by deforming nearest real energy levels into complex ones through an EP, which stimulates an intuitive understanding by projecting the Hamiltonian into two-level subspaces. The model in Eq.~\eqref{eq:OBCmodel1} is inversion symmetric, so that its eigenstates can be categorized with even or odd parity. Under OBC, accidental degeneracy in the spectrum [Fig.~\ref{fig:spectrum} (a)] can exist for modes with opposite parity \cite{supplemental}. In a subspace effective theory, the perturbation $V$ can only couple modes belonging to opposite parities, as it is inversion anti-symmetric, i.e., $\mathcal PV\mathcal P=-V$, with $\mathcal P$ being the inversion operator; hence, the perturbative terms in the effective Hamiltonian are off-diagonal and anti-Hermitian. The effective Hamiltonian for two eigenmodes involved in a $PT$ breaking takes the form
\begin{eqnarray}
H_{\text{eff}}=\Delta_{12}\sigma_z+ig(d_x\sigma_x+ d_y\sigma_y),
\end{eqnarray}
where $\Delta_{12}$ represents the energy gap, and  $d_x, d_y$ are functions of the involved modes and the perturbation. At the degenerate point $\Delta_{12}=0$, the eigenvalue of the effective Hamiltonian is $\pm i g\sqrt{d_x^2+d_y^2}$, namely, the $PT$ symmetry is broken if $g\neq 0$, agreeing with the vanishing threshold in Fig.~\ref{fig:PTbreaking_OBC} (a) (in the $t_2>t_1/4$ regime). Here, one should notice that the vanishing $PT$ breaking threshold is attributed to the inversion anti-symmetry of perturbation, which can couple equal-energy modes of opposite parity. When an inversion symmetric non-Hermitian perturbation is considered, it would mix gapped modes belonging to the same parity. The perturbation needs to be strong enough to overcome the gap, which corresponds to a finite threshold \cite{supplemental}.

\textit{Conclusions}. We show that a local non-Hermitian perturbation in Hermitian lattices can generically induce scale-free localization for continuous-spectrum eigenstates. Furthermore, the same mechanism can generate a collection of exponentially localized modes when the local non-Hermitian perturbation sits at a finite distance to the boundary, and the number of these modes is proportional to the distance. When $PT$ symmetry is present, we show that the scale-free localization emerges simultaneously with the $PT$ breaking. The continuous-spectrum $PT$ breaking (associated with scale-free modes) arising from local boundary perturbation is dictated by a concise criterion based on band structure.




One of the most promising platforms for verifying our results is optics systems, where gain and loss can be conveniently controlled, based on which tremendous progress on $PT$ symmetric non-Hermitian physics has been achieved~\cite{makris2008beam,El-Ganainy2007,Ruter2010,guo2009complex, Zhen2015,Ding2015,Cerjan2016}. Another promising setup is an array of cavities, which plays important roles in studying driven-dissipative quantum systems~\cite{Carusotto2009,Umucal2012,Bardyn2012, Dartiailh2017,Deuar2021}. Its inherent non-Hermiticity can be engineered according to our proposals. Our results are also relevant to other boundary-driven/dissipated systems. As a final remark, it will be interesting to generalize our results to many-body systems; for example, the XXZ spin chain subject to a $PT$ symmetric imaginary magnetic field on boundaries for which integrable techniques based on the Bethe ansatz are available ~\cite{FCAlcaraz1987,Wang2023Scale}.
%

\textit{Acknowledgements}. This work is supported by NSFC under Grant No. 12125405. 

\bibliography{Floquet,dirac}

\begin{thebibliography}{55}%
\makeatletter
\providecommand \@ifxundefined [1]{%
 \@ifx{#1\undefined}
}%
\providecommand \@ifnum [1]{%
 \ifnum #1\expandafter \@firstoftwo
 \else \expandafter \@secondoftwo
 \fi
}%
\providecommand \@ifx [1]{%
 \ifx #1\expandafter \@firstoftwo
 \else \expandafter \@secondoftwo
 \fi
}%
\providecommand \natexlab [1]{#1}%
\providecommand \enquote  [1]{``#1''}%
\providecommand \bibnamefont  [1]{#1}%
\providecommand \bibfnamefont [1]{#1}%
\providecommand \citenamefont [1]{#1}%
\providecommand \href@noop [0]{\@secondoftwo}%
\providecommand \href [0]{\begingroup \@sanitize@url \@href}%
\providecommand \@href[1]{\@@startlink{#1}\@@href}%
\providecommand \@@href[1]{\endgroup#1\@@endlink}%
\providecommand \@sanitize@url [0]{\catcode `\\12\catcode `\$12\catcode
  `\&12\catcode `\#12\catcode `\^12\catcode `\_12\catcode `\%12\relax}%
\providecommand \@@startlink[1]{}%
\providecommand \@@endlink[0]{}%
\providecommand \url  [0]{\begingroup\@sanitize@url \@url }%
\providecommand \@url [1]{\endgroup\@href {#1}{\urlprefix }}%
\providecommand \urlprefix  [0]{URL }%
\providecommand \Eprint [0]{\href }%
\providecommand \doibase [0]{http://dx.doi.org/}%
\providecommand \selectlanguage [0]{\@gobble}%
\providecommand \bibinfo  [0]{\@secondoftwo}%
\providecommand \bibfield  [0]{\@secondoftwo}%
\providecommand \translation [1]{[#1]}%
\providecommand \BibitemOpen [0]{}%
\providecommand \bibitemStop [0]{}%
\providecommand \bibitemNoStop [0]{.\EOS\space}%
\providecommand \EOS [0]{\spacefactor3000\relax}%
\providecommand \BibitemShut  [1]{\csname bibitem#1\endcsname}%
\let\auto@bib@innerbib\@empty
\bibitem [{\citenamefont {Ashida}\ \emph {et~al.}(2020)\citenamefont {Ashida},
  \citenamefont {Gong},\ and\ \citenamefont {Ueda}}]{Ashida2020}%
  \BibitemOpen
  \bibfield  {author} {\bibinfo {author} {\bibfnamefont {Yuto}\ \bibnamefont
  {Ashida}}, \bibinfo {author} {\bibfnamefont {Zongping}\ \bibnamefont {Gong}},
  \ and\ \bibinfo {author} {\bibfnamefont {Masahito}\ \bibnamefont {Ueda}},\
  }\bibfield  {title} {\enquote {\bibinfo {title} {Non-hermitian physics},}\
  }\href {\doibase 10.1080/00018732.2021.1876991} {\bibfield  {journal}
  {\bibinfo  {journal} {Advances in Physics}\ }\textbf {\bibinfo {volume}
  {69}},\ \bibinfo {pages} {249--435} (\bibinfo {year} {2020})}\BibitemShut
  {NoStop}%
\bibitem [{\citenamefont {Sternheim}\ and\ \citenamefont
  {Walker}(1972)}]{Sternheim1972}%
  \BibitemOpen
  \bibfield  {author} {\bibinfo {author} {\bibfnamefont {Morton~M.}\
  \bibnamefont {Sternheim}}\ and\ \bibinfo {author} {\bibfnamefont {James~F.}\
  \bibnamefont {Walker}},\ }\bibfield  {title} {\enquote {\bibinfo {title}
  {Non-hermitian hamiltonians, decaying states, and perturbation theory},}\
  }\href {\doibase 10.1103/PhysRevC.6.114} {\bibfield  {journal} {\bibinfo
  {journal} {Phys. Rev. C}\ }\textbf {\bibinfo {volume} {6}},\ \bibinfo {pages}
  {114--121} (\bibinfo {year} {1972})}\BibitemShut {NoStop}%
\bibitem [{\citenamefont {Lin}\ \emph {et~al.}(2011)\citenamefont {Lin},
  \citenamefont {Ramezani}, \citenamefont {Eichelkraut}, \citenamefont
  {Kottos}, \citenamefont {Cao},\ and\ \citenamefont
  {Christodoulides}}]{lin2011unidirectional}%
  \BibitemOpen
  \bibfield  {author} {\bibinfo {author} {\bibfnamefont {Zin}\ \bibnamefont
  {Lin}}, \bibinfo {author} {\bibfnamefont {Hamidreza}\ \bibnamefont
  {Ramezani}}, \bibinfo {author} {\bibfnamefont {Toni}\ \bibnamefont
  {Eichelkraut}}, \bibinfo {author} {\bibfnamefont {Tsampikos}\ \bibnamefont
  {Kottos}}, \bibinfo {author} {\bibfnamefont {Hui}\ \bibnamefont {Cao}}, \
  and\ \bibinfo {author} {\bibfnamefont {Demetrios~N.}\ \bibnamefont
  {Christodoulides}},\ }\bibfield  {title} {\enquote {\bibinfo {title}
  {Unidirectional invisibility induced by $\mathcal{P}\mathcal{T}$-symmetric
  periodic structures},}\ }\href {\doibase 10.1103/PhysRevLett.106.213901}
  {\bibfield  {journal} {\bibinfo  {journal} {Phys. Rev. Lett.}\ }\textbf
  {\bibinfo {volume} {106}},\ \bibinfo {pages} {213901} (\bibinfo {year}
  {2011})}\BibitemShut {NoStop}%
\bibitem [{\citenamefont {Feng}\ \emph {et~al.}(2013)\citenamefont {Feng},
  \citenamefont {Xu}, \citenamefont {Fegadolli}, \citenamefont {Lu},
  \citenamefont {Oliveira}, \citenamefont {Almeida}, \citenamefont {Chen},\
  and\ \citenamefont {Scherer}}]{Feng2013}%
  \BibitemOpen
  \bibfield  {author} {\bibinfo {author} {\bibfnamefont {Liang}\ \bibnamefont
  {Feng}}, \bibinfo {author} {\bibfnamefont {Ye-Long}\ \bibnamefont {Xu}},
  \bibinfo {author} {\bibfnamefont {William~S.}\ \bibnamefont {Fegadolli}},
  \bibinfo {author} {\bibfnamefont {Ming-Hui}\ \bibnamefont {Lu}}, \bibinfo
  {author} {\bibfnamefont {Jos{\'e} E.~B.}\ \bibnamefont {Oliveira}}, \bibinfo
  {author} {\bibfnamefont {Vilson~R.}\ \bibnamefont {Almeida}}, \bibinfo
  {author} {\bibfnamefont {Yan-Feng}\ \bibnamefont {Chen}}, \ and\ \bibinfo
  {author} {\bibfnamefont {Axel}\ \bibnamefont {Scherer}},\ }\bibfield  {title}
  {\enquote {\bibinfo {title} {Experimental demonstration of a unidirectional
  reflectionless parity-time metamaterial at optical frequencies},}\ }\href
  {\doibase 10.1038/nmat3495} {\bibfield  {journal} {\bibinfo  {journal}
  {Nature Materials}\ }\textbf {\bibinfo {volume} {12}},\ \bibinfo {pages}
  {108--113} (\bibinfo {year} {2013})}\BibitemShut {NoStop}%
\bibitem [{\citenamefont {Hodaei}\ \emph {et~al.}(2014)\citenamefont {Hodaei},
  \citenamefont {Miri}, \citenamefont {Heinrich}, \citenamefont
  {Christodoulides},\ and\ \citenamefont {Khajavikhan}}]{hodaei2014PT}%
  \BibitemOpen
  \bibfield  {author} {\bibinfo {author} {\bibfnamefont {Hossein}\ \bibnamefont
  {Hodaei}}, \bibinfo {author} {\bibfnamefont {Mohammad-Ali}\ \bibnamefont
  {Miri}}, \bibinfo {author} {\bibfnamefont {Matthias}\ \bibnamefont
  {Heinrich}}, \bibinfo {author} {\bibfnamefont {Demetrios~N}\ \bibnamefont
  {Christodoulides}}, \ and\ \bibinfo {author} {\bibfnamefont {Mercedeh}\
  \bibnamefont {Khajavikhan}},\ }\bibfield  {title} {\enquote {\bibinfo {title}
  {Parity-time--symmetric microring lasers},}\ }\href@noop {} {\bibfield
  {journal} {\bibinfo  {journal} {Science}\ }\textbf {\bibinfo {volume}
  {346}},\ \bibinfo {pages} {975--978} (\bibinfo {year} {2014})}\BibitemShut
  {NoStop}%
\bibitem [{\citenamefont {Feng}\ \emph {et~al.}(2014)\citenamefont {Feng},
  \citenamefont {Wong}, \citenamefont {Ma}, \citenamefont {Wang},\ and\
  \citenamefont {Zhang}}]{feng2014singlemode}%
  \BibitemOpen
  \bibfield  {author} {\bibinfo {author} {\bibfnamefont {Liang}\ \bibnamefont
  {Feng}}, \bibinfo {author} {\bibfnamefont {Zi~Jing}\ \bibnamefont {Wong}},
  \bibinfo {author} {\bibfnamefont {Ren-Min}\ \bibnamefont {Ma}}, \bibinfo
  {author} {\bibfnamefont {Yuan}\ \bibnamefont {Wang}}, \ and\ \bibinfo
  {author} {\bibfnamefont {Xiang}\ \bibnamefont {Zhang}},\ }\bibfield  {title}
  {\enquote {\bibinfo {title} {Single-mode laser by parity-time symmetry
  breaking},}\ }\href@noop {} {\bibfield  {journal} {\bibinfo  {journal}
  {Science}\ }\textbf {\bibinfo {volume} {346}},\ \bibinfo {pages} {972--975}
  (\bibinfo {year} {2014})}\BibitemShut {NoStop}%
\bibitem [{\citenamefont {Hodaei}\ \emph {et~al.}(2017)\citenamefont {Hodaei},
  \citenamefont {Hassan}, \citenamefont {Wittek}, \citenamefont
  {Garcia-Gracia}, \citenamefont {El-Ganainy}, \citenamefont
  {Christodoulides},\ and\ \citenamefont {Khajavikhan}}]{Hodaei2017}%
  \BibitemOpen
  \bibfield  {author} {\bibinfo {author} {\bibfnamefont {Hossein}\ \bibnamefont
  {Hodaei}}, \bibinfo {author} {\bibfnamefont {Absar~U.}\ \bibnamefont
  {Hassan}}, \bibinfo {author} {\bibfnamefont {Steffen}\ \bibnamefont
  {Wittek}}, \bibinfo {author} {\bibfnamefont {Hipolito}\ \bibnamefont
  {Garcia-Gracia}}, \bibinfo {author} {\bibfnamefont {Ramy}\ \bibnamefont
  {El-Ganainy}}, \bibinfo {author} {\bibfnamefont {Demetrios~N.}\ \bibnamefont
  {Christodoulides}}, \ and\ \bibinfo {author} {\bibfnamefont {Mercedeh}\
  \bibnamefont {Khajavikhan}},\ }\bibfield  {title} {\enquote {\bibinfo {title}
  {Enhanced sensitivity at higher-order exceptional points},}\ }\href {\doibase
  10.1038/nature23280} {\bibfield  {journal} {\bibinfo  {journal} {Nature}\
  }\textbf {\bibinfo {volume} {548}},\ \bibinfo {pages} {187--191} (\bibinfo
  {year} {2017})}\BibitemShut {NoStop}%
\bibitem [{\citenamefont {Chen}\ and\ \citenamefont {Jung}(2016)}]{Chen2016}%
  \BibitemOpen
  \bibfield  {author} {\bibinfo {author} {\bibfnamefont {Pai-Yen}\ \bibnamefont
  {Chen}}\ and\ \bibinfo {author} {\bibfnamefont {Jeil}\ \bibnamefont {Jung}},\
  }\bibfield  {title} {\enquote {\bibinfo {title} {$\mathcal{P}\mathcal{T}$
  symmetry and singularity-enhanced sensing based on photoexcited graphene
  metasurfaces},}\ }\href {\doibase 10.1103/PhysRevApplied.5.064018} {\bibfield
   {journal} {\bibinfo  {journal} {Phys. Rev. Applied}\ }\textbf {\bibinfo
  {volume} {5}},\ \bibinfo {pages} {064018} (\bibinfo {year}
  {2016})}\BibitemShut {NoStop}%
\bibitem [{\citenamefont {Liu}\ \emph {et~al.}(2016)\citenamefont {Liu},
  \citenamefont {Zhang}, \citenamefont {\"Ozdemir}, \citenamefont {Peng},
  \citenamefont {Jing}, \citenamefont {L\"u}, \citenamefont {Li}, \citenamefont
  {Yang}, \citenamefont {Nori},\ and\ \citenamefont {Liu}}]{Liu2016}%
  \BibitemOpen
  \bibfield  {author} {\bibinfo {author} {\bibfnamefont {Zhong-Peng}\
  \bibnamefont {Liu}}, \bibinfo {author} {\bibfnamefont {Jing}\ \bibnamefont
  {Zhang}}, \bibinfo {author} {\bibfnamefont {\ifmmode \mbox{\c{S}}\else
  \c{S}\fi{}ahin~Kaya}\ \bibnamefont {\"Ozdemir}}, \bibinfo {author}
  {\bibfnamefont {Bo}~\bibnamefont {Peng}}, \bibinfo {author} {\bibfnamefont
  {Hui}\ \bibnamefont {Jing}}, \bibinfo {author} {\bibfnamefont {Xin-You}\
  \bibnamefont {L\"u}}, \bibinfo {author} {\bibfnamefont {Chun-Wen}\
  \bibnamefont {Li}}, \bibinfo {author} {\bibfnamefont {Lan}\ \bibnamefont
  {Yang}}, \bibinfo {author} {\bibfnamefont {Franco}\ \bibnamefont {Nori}}, \
  and\ \bibinfo {author} {\bibfnamefont {Yu-xi}\ \bibnamefont {Liu}},\
  }\bibfield  {title} {\enquote {\bibinfo {title} {Metrology with
  $\mathcal{PT}$-symmetric cavities: Enhanced sensitivity near the
  $\mathcal{PT}$-phase transition},}\ }\href {\doibase
  10.1103/PhysRevLett.117.110802} {\bibfield  {journal} {\bibinfo  {journal}
  {Phys. Rev. Lett.}\ }\textbf {\bibinfo {volume} {117}},\ \bibinfo {pages}
  {110802} (\bibinfo {year} {2016})}\BibitemShut {NoStop}%
\bibitem [{\citenamefont {Chen}\ \emph {et~al.}(2018)\citenamefont {Chen},
  \citenamefont {Sakhdari}, \citenamefont {Hajizadegan}, \citenamefont {Cui},
  \citenamefont {Cheng}, \citenamefont {El-Ganainy},\ and\ \citenamefont
  {Al{\`u}}}]{Chen2018}%
  \BibitemOpen
  \bibfield  {author} {\bibinfo {author} {\bibfnamefont {Pai-Yen}\ \bibnamefont
  {Chen}}, \bibinfo {author} {\bibfnamefont {Maryam}\ \bibnamefont {Sakhdari}},
  \bibinfo {author} {\bibfnamefont {Mehdi}\ \bibnamefont {Hajizadegan}},
  \bibinfo {author} {\bibfnamefont {Qingsong}\ \bibnamefont {Cui}}, \bibinfo
  {author} {\bibfnamefont {Mark Ming-Cheng}\ \bibnamefont {Cheng}}, \bibinfo
  {author} {\bibfnamefont {Ramy}\ \bibnamefont {El-Ganainy}}, \ and\ \bibinfo
  {author} {\bibfnamefont {Andrea}\ \bibnamefont {Al{\`u}}},\ }\bibfield
  {title} {\enquote {\bibinfo {title} {Generalized parity--time symmetry
  condition for enhanced sensor telemetry},}\ }\href {\doibase
  10.1038/s41928-018-0072-6} {\bibfield  {journal} {\bibinfo  {journal} {Nature
  Electronics}\ }\textbf {\bibinfo {volume} {1}},\ \bibinfo {pages} {297--304}
  (\bibinfo {year} {2018})}\BibitemShut {NoStop}%
\bibitem [{\citenamefont {Bender}\ and\ \citenamefont
  {Boettcher}(1998)}]{bender1998real}%
  \BibitemOpen
  \bibfield  {author} {\bibinfo {author} {\bibfnamefont {Carl~M.}\ \bibnamefont
  {Bender}}\ and\ \bibinfo {author} {\bibfnamefont {Stefan}\ \bibnamefont
  {Boettcher}},\ }\bibfield  {title} {\enquote {\bibinfo {title} {Real spectra
  in non-hermitian hamiltonians having pt symmetry},}\ }\href {\doibase
  10.1103/PhysRevLett.80.5243} {\bibfield  {journal} {\bibinfo  {journal}
  {Phys. Rev. Lett.}\ }\textbf {\bibinfo {volume} {80}},\ \bibinfo {pages}
  {5243--5246} (\bibinfo {year} {1998})}\BibitemShut {NoStop}%
\bibitem [{\citenamefont {Bender}(2007)}]{Bender_2007}%
  \BibitemOpen
  \bibfield  {author} {\bibinfo {author} {\bibfnamefont {Carl~M}\ \bibnamefont
  {Bender}},\ }\bibfield  {title} {\enquote {\bibinfo {title} {Making sense of
  non-hermitian hamiltonians},}\ }\href {\doibase 10.1088/0034-4885/70/6/r03}
  {\bibfield  {journal} {\bibinfo  {journal} {Reports on Progress in Physics}\
  }\textbf {\bibinfo {volume} {70}},\ \bibinfo {pages} {947--1018} (\bibinfo
  {year} {2007})}\BibitemShut {NoStop}%
\bibitem [{\citenamefont {{\"O}zdemir}\ \emph {et~al.}(2019)\citenamefont
  {{\"O}zdemir}, \citenamefont {Rotter}, \citenamefont {Nori},\ and\
  \citenamefont {Yang}}]{Ozdemir2019}%
  \BibitemOpen
  \bibfield  {author} {\bibinfo {author} {\bibfnamefont {{\c{S}}~K.}\
  \bibnamefont {{\"O}zdemir}}, \bibinfo {author} {\bibfnamefont
  {S.}~\bibnamefont {Rotter}}, \bibinfo {author} {\bibfnamefont
  {F.}~\bibnamefont {Nori}}, \ and\ \bibinfo {author} {\bibfnamefont
  {L.}~\bibnamefont {Yang}},\ }\bibfield  {title} {\enquote {\bibinfo {title}
  {Parity--time symmetry and exceptional points in photonics},}\ }\href
  {\doibase 10.1038/s41563-019-0304-9} {\bibfield  {journal} {\bibinfo
  {journal} {Nature Materials}\ }\textbf {\bibinfo {volume} {18}},\ \bibinfo
  {pages} {783--798} (\bibinfo {year} {2019})}\BibitemShut {NoStop}%
\bibitem [{\citenamefont {El-Ganainy}\ \emph {et~al.}(2018)\citenamefont
  {El-Ganainy}, \citenamefont {Makris}, \citenamefont {Khajavikhan},
  \citenamefont {Musslimani}, \citenamefont {Rotter},\ and\ \citenamefont
  {Christodoulides}}]{El-Ganainy2018}%
  \BibitemOpen
  \bibfield  {author} {\bibinfo {author} {\bibfnamefont {Ramy}\ \bibnamefont
  {El-Ganainy}}, \bibinfo {author} {\bibfnamefont {Konstantinos~G.}\
  \bibnamefont {Makris}}, \bibinfo {author} {\bibfnamefont {Mercedeh}\
  \bibnamefont {Khajavikhan}}, \bibinfo {author} {\bibfnamefont {Ziad~H.}\
  \bibnamefont {Musslimani}}, \bibinfo {author} {\bibfnamefont {Stefan}\
  \bibnamefont {Rotter}}, \ and\ \bibinfo {author} {\bibfnamefont
  {Demetrios~N.}\ \bibnamefont {Christodoulides}},\ }\bibfield  {title}
  {\enquote {\bibinfo {title} {Non-hermitian physics and pt symmetry},}\ }\href
  {\doibase 10.1038/nphys4323} {\bibfield  {journal} {\bibinfo  {journal}
  {Nature Physics}\ }\textbf {\bibinfo {volume} {14}},\ \bibinfo {pages}
  {11--19} (\bibinfo {year} {2018})}\BibitemShut {NoStop}%
\bibitem [{\citenamefont {Miri}\ and\ \citenamefont {Alù}(2019)}]{Miri2019EP}%
  \BibitemOpen
  \bibfield  {author} {\bibinfo {author} {\bibfnamefont {Mohammad-Ali}\
  \bibnamefont {Miri}}\ and\ \bibinfo {author} {\bibfnamefont {Andrea}\
  \bibnamefont {Alù}},\ }\bibfield  {title} {\enquote {\bibinfo {title}
  {Exceptional points in optics and photonics},}\ }\href {\doibase
  10.1126/science.aar7709} {\bibfield  {journal} {\bibinfo  {journal}
  {Science}\ }\textbf {\bibinfo {volume} {363}},\ \bibinfo {pages} {eaar7709}
  (\bibinfo {year} {2019})}\BibitemShut {NoStop}%
\bibitem [{\citenamefont {Yao}\ and\ \citenamefont {Wang}(2018)}]{yao2018edge}%
  \BibitemOpen
  \bibfield  {author} {\bibinfo {author} {\bibfnamefont {Shunyu}\ \bibnamefont
  {Yao}}\ and\ \bibinfo {author} {\bibfnamefont {Zhong}\ \bibnamefont {Wang}},\
  }\bibfield  {title} {\enquote {\bibinfo {title} {Edge states and topological
  invariants of non-hermitian systems},}\ }\href {\doibase
  10.1103/PhysRevLett.121.086803} {\bibfield  {journal} {\bibinfo  {journal}
  {Phys. Rev. Lett.}\ }\textbf {\bibinfo {volume} {121}},\ \bibinfo {pages}
  {086803} (\bibinfo {year} {2018})}\BibitemShut {NoStop}%
\bibitem [{\citenamefont {Yao}\ \emph {et~al.}(2018)\citenamefont {Yao},
  \citenamefont {Song},\ and\ \citenamefont {Wang}}]{yao2018chern}%
  \BibitemOpen
  \bibfield  {author} {\bibinfo {author} {\bibfnamefont {Shunyu}\ \bibnamefont
  {Yao}}, \bibinfo {author} {\bibfnamefont {Fei}\ \bibnamefont {Song}}, \ and\
  \bibinfo {author} {\bibfnamefont {Zhong}\ \bibnamefont {Wang}},\ }\bibfield
  {title} {\enquote {\bibinfo {title} {Non-hermitian chern bands},}\ }\href
  {\doibase 10.1103/PhysRevLett.121.136802} {\bibfield  {journal} {\bibinfo
  {journal} {Phys. Rev. Lett.}\ }\textbf {\bibinfo {volume} {121}},\ \bibinfo
  {pages} {136802} (\bibinfo {year} {2018})}\BibitemShut {NoStop}%
\bibitem [{\citenamefont {Kunst}\ \emph {et~al.}(2018)\citenamefont {Kunst},
  \citenamefont {Edvardsson}, \citenamefont {Budich},\ and\ \citenamefont
  {Bergholtz}}]{kunst2018biorthogonal}%
  \BibitemOpen
  \bibfield  {author} {\bibinfo {author} {\bibfnamefont {Flore~K.}\
  \bibnamefont {Kunst}}, \bibinfo {author} {\bibfnamefont {Elisabet}\
  \bibnamefont {Edvardsson}}, \bibinfo {author} {\bibfnamefont {Jan~Carl}\
  \bibnamefont {Budich}}, \ and\ \bibinfo {author} {\bibfnamefont {Emil~J.}\
  \bibnamefont {Bergholtz}},\ }\bibfield  {title} {\enquote {\bibinfo {title}
  {Biorthogonal bulk-boundary correspondence in non-hermitian systems},}\
  }\href {\doibase 10.1103/PhysRevLett.121.026808} {\bibfield  {journal}
  {\bibinfo  {journal} {Phys. Rev. Lett.}\ }\textbf {\bibinfo {volume} {121}},\
  \bibinfo {pages} {026808} (\bibinfo {year} {2018})}\BibitemShut {NoStop}%
\bibitem [{\citenamefont {Lee}\ and\ \citenamefont
  {Thomale}(2019)}]{lee2018anatomy}%
  \BibitemOpen
  \bibfield  {author} {\bibinfo {author} {\bibfnamefont {Ching~Hua}\
  \bibnamefont {Lee}}\ and\ \bibinfo {author} {\bibfnamefont {Ronny}\
  \bibnamefont {Thomale}},\ }\bibfield  {title} {\enquote {\bibinfo {title}
  {Anatomy of skin modes and topology in non-hermitian systems},}\ }\href
  {\doibase 10.1103/PhysRevB.99.201103} {\bibfield  {journal} {\bibinfo
  {journal} {Phys. Rev. B}\ }\textbf {\bibinfo {volume} {99}},\ \bibinfo
  {pages} {201103} (\bibinfo {year} {2019})}\BibitemShut {NoStop}%
\bibitem [{\citenamefont {Longhi}(2019)}]{Longhi2019Probing}%
  \BibitemOpen
  \bibfield  {author} {\bibinfo {author} {\bibfnamefont {Stefano}\ \bibnamefont
  {Longhi}},\ }\bibfield  {title} {\enquote {\bibinfo {title} {Probing
  non-hermitian skin effect and non-bloch phase transitions},}\ }\href
  {\doibase 10.1103/PhysRevResearch.1.023013} {\bibfield  {journal} {\bibinfo
  {journal} {Phys. Rev. Research}\ }\textbf {\bibinfo {volume} {1}},\ \bibinfo
  {pages} {023013} (\bibinfo {year} {2019})}\BibitemShut {NoStop}%
\bibitem [{\citenamefont {Helbig}\ \emph {et~al.}(2020)\citenamefont {Helbig},
  \citenamefont {Hofmann}, \citenamefont {Imhof}, \citenamefont {Abdelghany},
  \citenamefont {Kiessling}, \citenamefont {Molenkamp}, \citenamefont {Lee},
  \citenamefont {Szameit}, \citenamefont {Greiter},\ and\ \citenamefont
  {Thomale}}]{Helbig2019NHSE}%
  \BibitemOpen
  \bibfield  {author} {\bibinfo {author} {\bibfnamefont {T.}~\bibnamefont
  {Helbig}}, \bibinfo {author} {\bibfnamefont {T.}~\bibnamefont {Hofmann}},
  \bibinfo {author} {\bibfnamefont {S.}~\bibnamefont {Imhof}}, \bibinfo
  {author} {\bibfnamefont {M.}~\bibnamefont {Abdelghany}}, \bibinfo {author}
  {\bibfnamefont {T.}~\bibnamefont {Kiessling}}, \bibinfo {author}
  {\bibfnamefont {L.~W.}\ \bibnamefont {Molenkamp}}, \bibinfo {author}
  {\bibfnamefont {C.~H.}\ \bibnamefont {Lee}}, \bibinfo {author} {\bibfnamefont
  {A.}~\bibnamefont {Szameit}}, \bibinfo {author} {\bibfnamefont
  {M.}~\bibnamefont {Greiter}}, \ and\ \bibinfo {author} {\bibfnamefont
  {R.}~\bibnamefont {Thomale}},\ }\bibfield  {title} {\enquote {\bibinfo
  {title} {Generalized bulk--boundary correspondence in non-hermitian
  topolectrical circuits},}\ }\href@noop {} {\bibfield  {journal} {\bibinfo
  {journal} {Nature Physics}\ }\textbf {\bibinfo {volume} {16}},\ \bibinfo
  {pages} {747} (\bibinfo {year} {2020})}\BibitemShut {NoStop}%
\bibitem [{\citenamefont {Xiao}\ \emph {et~al.}(2020)\citenamefont {Xiao},
  \citenamefont {Deng}, \citenamefont {Wang}, \citenamefont {Zhu},
  \citenamefont {Wang}, \citenamefont {Yi},\ and\ \citenamefont
  {Xue}}]{xiao2020non}%
  \BibitemOpen
  \bibfield  {author} {\bibinfo {author} {\bibfnamefont {Lei}\ \bibnamefont
  {Xiao}}, \bibinfo {author} {\bibfnamefont {Tianshu}\ \bibnamefont {Deng}},
  \bibinfo {author} {\bibfnamefont {Kunkun}\ \bibnamefont {Wang}}, \bibinfo
  {author} {\bibfnamefont {Gaoyan}\ \bibnamefont {Zhu}}, \bibinfo {author}
  {\bibfnamefont {Zhong}\ \bibnamefont {Wang}}, \bibinfo {author}
  {\bibfnamefont {Wei}\ \bibnamefont {Yi}}, \ and\ \bibinfo {author}
  {\bibfnamefont {Peng}\ \bibnamefont {Xue}},\ }\bibfield  {title} {\enquote
  {\bibinfo {title} {Non-hermitian bulk--boundary correspondence in quantum
  dynamics},}\ }\href@noop {} {\bibfield  {journal} {\bibinfo  {journal}
  {Nature Physics}\ ,\ \bibinfo {pages} {1--6}} (\bibinfo {year}
  {2020})}\BibitemShut {NoStop}%
\bibitem [{\citenamefont {Martinez~Alvarez}\ \emph {et~al.}(2018)\citenamefont
  {Martinez~Alvarez}, \citenamefont {Barrios~Vargas},\ and\ \citenamefont
  {Foa~Torres}}]{alvarez2017}%
  \BibitemOpen
  \bibfield  {author} {\bibinfo {author} {\bibfnamefont {V.~M.}\ \bibnamefont
  {Martinez~Alvarez}}, \bibinfo {author} {\bibfnamefont {J.~E.}\ \bibnamefont
  {Barrios~Vargas}}, \ and\ \bibinfo {author} {\bibfnamefont {L.~E.~F.}\
  \bibnamefont {Foa~Torres}},\ }\bibfield  {title} {\enquote {\bibinfo {title}
  {Non-hermitian robust edge states in one dimension: Anomalous localization
  and eigenspace condensation at exceptional points},}\ }\href {\doibase
  10.1103/PhysRevB.97.121401} {\bibfield  {journal} {\bibinfo  {journal} {Phys.
  Rev. B}\ }\textbf {\bibinfo {volume} {97}},\ \bibinfo {pages} {121401}
  (\bibinfo {year} {2018})}\BibitemShut {NoStop}%
\bibitem [{\citenamefont {Ghatak}\ \emph {et~al.}(2020)\citenamefont {Ghatak},
  \citenamefont {Brandenbourger}, \citenamefont {van Wezel},\ and\
  \citenamefont {Coulais}}]{Ghatak2019NHSE}%
  \BibitemOpen
  \bibfield  {author} {\bibinfo {author} {\bibfnamefont {Ananya}\ \bibnamefont
  {Ghatak}}, \bibinfo {author} {\bibfnamefont {Martin}\ \bibnamefont
  {Brandenbourger}}, \bibinfo {author} {\bibfnamefont {Jasper}\ \bibnamefont
  {van Wezel}}, \ and\ \bibinfo {author} {\bibfnamefont {Corentin}\
  \bibnamefont {Coulais}},\ }\bibfield  {title} {\enquote {\bibinfo {title}
  {Observation of non-hermitian topology and its bulk--edge correspondence in
  an active mechanical metamaterial},}\ }\href@noop {} {\bibfield  {journal}
  {\bibinfo  {journal} {Proceedings of the National Academy of Sciences}\
  }\textbf {\bibinfo {volume} {117}},\ \bibinfo {pages} {29561--29568}
  (\bibinfo {year} {2020})}\BibitemShut {NoStop}%
\bibitem [{\citenamefont {{Wang}}\ \emph {et~al.}(2022)\citenamefont {{Wang}},
  \citenamefont {{Wang}},\ and\ \citenamefont {{Ma}}}]{Wang2022morphing}%
  \BibitemOpen
  \bibfield  {author} {\bibinfo {author} {\bibfnamefont {Wei}\ \bibnamefont
  {{Wang}}}, \bibinfo {author} {\bibfnamefont {Xulong}\ \bibnamefont {{Wang}}},
  \ and\ \bibinfo {author} {\bibfnamefont {Guancong}\ \bibnamefont {{Ma}}},\
  }\bibfield  {title} {\enquote {\bibinfo {title} {{Non-Hermitian morphing of
  topological modes}},}\ }\href {\doibase 10.1038/s41586-022-04929-1}
  {\bibfield  {journal} {\bibinfo  {journal} {Nature}\ }\textbf {\bibinfo
  {volume} {608}},\ \bibinfo {pages} {50--55} (\bibinfo {year} {2022})},\
  \Eprint {http://arxiv.org/abs/2203.02147} {arXiv:2203.02147 [physics.app-ph]}
  \BibitemShut {NoStop}%
\bibitem [{\citenamefont {Bergholtz}\ \emph {et~al.}(2021)\citenamefont
  {Bergholtz}, \citenamefont {Budich},\ and\ \citenamefont
  {Kunst}}]{Bergholtz2021RMP}%
  \BibitemOpen
  \bibfield  {author} {\bibinfo {author} {\bibfnamefont {Emil~J.}\ \bibnamefont
  {Bergholtz}}, \bibinfo {author} {\bibfnamefont {Jan~Carl}\ \bibnamefont
  {Budich}}, \ and\ \bibinfo {author} {\bibfnamefont {Flore~K.}\ \bibnamefont
  {Kunst}},\ }\bibfield  {title} {\enquote {\bibinfo {title} {Exceptional
  topology of non-hermitian systems},}\ }\href {\doibase
  10.1103/RevModPhys.93.015005} {\bibfield  {journal} {\bibinfo  {journal}
  {Rev. Mod. Phys.}\ }\textbf {\bibinfo {volume} {93}},\ \bibinfo {pages}
  {015005} (\bibinfo {year} {2021})}\BibitemShut {NoStop}%
\bibitem [{\citenamefont {Landi}\ \emph {et~al.}(2021)\citenamefont {Landi},
  \citenamefont {Poletti},\ and\ \citenamefont {Schaller}}]{Landi2021}%
  \BibitemOpen
  \bibfield  {author} {\bibinfo {author} {\bibfnamefont {Gabriel~T.}\
  \bibnamefont {Landi}}, \bibinfo {author} {\bibfnamefont {Dario}\ \bibnamefont
  {Poletti}}, \ and\ \bibinfo {author} {\bibfnamefont {Gernot}\ \bibnamefont
  {Schaller}},\ }\bibfield  {title} {\enquote {\bibinfo {title}
  {Non-equilibrium boundary driven quantum systems: models, methods and
  properties},}\ }\href {\doibase 10.48550/ARXIV.2104.14350} {\  (\bibinfo
  {year} {2021}),\ 10.48550/ARXIV.2104.14350}\BibitemShut {NoStop}%
\bibitem [{\citenamefont {{Roccati}}\ \emph {et~al.}(2022)\citenamefont
  {{Roccati}}, \citenamefont {{Palma}}, \citenamefont {{Ciccarello}},\ and\
  \citenamefont {{Bagarello}}}]{Federico2022}%
  \BibitemOpen
  \bibfield  {author} {\bibinfo {author} {\bibfnamefont {Federico}\
  \bibnamefont {{Roccati}}}, \bibinfo {author} {\bibfnamefont {G.~Massimo}\
  \bibnamefont {{Palma}}}, \bibinfo {author} {\bibfnamefont {Francesco}\
  \bibnamefont {{Ciccarello}}}, \ and\ \bibinfo {author} {\bibfnamefont
  {Fabio}\ \bibnamefont {{Bagarello}}},\ }\bibfield  {title} {\enquote
  {\bibinfo {title} {{Non-Hermitian Physics and Master Equations}},}\ }\href
  {\doibase 10.1142/S1230161222500044} {\bibfield  {journal} {\bibinfo
  {journal} {Open Systems and Information Dynamics}\ }\textbf {\bibinfo
  {volume} {29}},\ \bibinfo {eid} {2250004} (\bibinfo {year} {2022})},\ \Eprint
  {http://arxiv.org/abs/2201.05367} {arXiv:2201.05367 [quant-ph]} \BibitemShut
  {NoStop}%
\bibitem [{\citenamefont {Wiersig}(2018)}]{Wiersig2018}%
  \BibitemOpen
  \bibfield  {author} {\bibinfo {author} {\bibfnamefont {Jan}\ \bibnamefont
  {Wiersig}},\ }\bibfield  {title} {\enquote {\bibinfo {title} {Role of
  nonorthogonality of energy eigenstates in quantum systems with localized
  losses},}\ }\href {\doibase 10.1103/PhysRevA.98.052105} {\bibfield  {journal}
  {\bibinfo  {journal} {Phys. Rev. A}\ }\textbf {\bibinfo {volume} {98}},\
  \bibinfo {pages} {052105} (\bibinfo {year} {2018})}\BibitemShut {NoStop}%
\bibitem [{\citenamefont {Fr\"oml}\ \emph {et~al.}(2020)\citenamefont
  {Fr\"oml}, \citenamefont {Muckel}, \citenamefont {Kollath}, \citenamefont
  {Chiocchetta},\ and\ \citenamefont {Diehl}}]{Heinrich2020}%
  \BibitemOpen
  \bibfield  {author} {\bibinfo {author} {\bibfnamefont {Heinrich}\
  \bibnamefont {Fr\"oml}}, \bibinfo {author} {\bibfnamefont {Christopher}\
  \bibnamefont {Muckel}}, \bibinfo {author} {\bibfnamefont {Corinna}\
  \bibnamefont {Kollath}}, \bibinfo {author} {\bibfnamefont {Alessio}\
  \bibnamefont {Chiocchetta}}, \ and\ \bibinfo {author} {\bibfnamefont
  {Sebastian}\ \bibnamefont {Diehl}},\ }\bibfield  {title} {\enquote {\bibinfo
  {title} {Ultracold quantum wires with localized losses: Many-body quantum
  zeno effect},}\ }\href {\doibase 10.1103/PhysRevB.101.144301} {\bibfield
  {journal} {\bibinfo  {journal} {Phys. Rev. B}\ }\textbf {\bibinfo {volume}
  {101}},\ \bibinfo {pages} {144301} (\bibinfo {year} {2020})}\BibitemShut
  {NoStop}%
\bibitem [{\citenamefont {Alba}\ and\ \citenamefont
  {Carollo}(2022)}]{Alba2022}%
  \BibitemOpen
  \bibfield  {author} {\bibinfo {author} {\bibfnamefont {Vincenzo}\
  \bibnamefont {Alba}}\ and\ \bibinfo {author} {\bibfnamefont {Federico}\
  \bibnamefont {Carollo}},\ }\bibfield  {title} {\enquote {\bibinfo {title}
  {Noninteracting fermionic systems with localized losses: Exact results in the
  hydrodynamic limit},}\ }\href {\doibase 10.1103/PhysRevB.105.054303}
  {\bibfield  {journal} {\bibinfo  {journal} {Phys. Rev. B}\ }\textbf {\bibinfo
  {volume} {105}},\ \bibinfo {pages} {054303} (\bibinfo {year}
  {2022})}\BibitemShut {NoStop}%
\bibitem [{\citenamefont {Datta}(1997)}]{datta1997electronic}%
  \BibitemOpen
  \bibfield  {author} {\bibinfo {author} {\bibfnamefont {Supriyo}\ \bibnamefont
  {Datta}},\ }\href@noop {} {\emph {\bibinfo {title} {Electronic transport in
  mesoscopic systems}}}\ (\bibinfo  {publisher} {Cambridge university press},\
  \bibinfo {year} {1997})\BibitemShut {NoStop}%
\bibitem [{\citenamefont {Burke}\ \emph {et~al.}(2020)\citenamefont {Burke},
  \citenamefont {Wiersig},\ and\ \citenamefont {Haque}}]{Burke2020localNH}%
  \BibitemOpen
  \bibfield  {author} {\bibinfo {author} {\bibfnamefont {Phillip~C.}\
  \bibnamefont {Burke}}, \bibinfo {author} {\bibfnamefont {Jan}\ \bibnamefont
  {Wiersig}}, \ and\ \bibinfo {author} {\bibfnamefont {Masudul}\ \bibnamefont
  {Haque}},\ }\bibfield  {title} {\enquote {\bibinfo {title} {Non-hermitian
  scattering on a tight-binding lattice},}\ }\href {\doibase
  10.1103/PhysRevA.102.012212} {\bibfield  {journal} {\bibinfo  {journal}
  {Phys. Rev. A}\ }\textbf {\bibinfo {volume} {102}},\ \bibinfo {pages}
  {012212} (\bibinfo {year} {2020})}\BibitemShut {NoStop}%
\bibitem [{\citenamefont {D\'ora}\ \emph {et~al.}(2021)\citenamefont {D\'ora},
  \citenamefont {Sticlet},\ and\ \citenamefont
  {Moca}}]{Dora2021Friedeloscillations}%
  \BibitemOpen
  \bibfield  {author} {\bibinfo {author} {\bibfnamefont {Bal\'azs}\
  \bibnamefont {D\'ora}}, \bibinfo {author} {\bibfnamefont {Doru}\ \bibnamefont
  {Sticlet}}, \ and\ \bibinfo {author} {\bibfnamefont {C\ifmmode \u{a}\else
  \u{a}\fi{}t\ifmmode \u{a}\else \u{a}\fi{}lin Pa\ifmmode
  \mbox{\c{s}}\else~\c{s}\fi{}cu}\ \bibnamefont {Moca}},\ }\bibfield  {title}
  {\enquote {\bibinfo {title} {Non-hermitian lindhard function and friedel
  oscillations},}\ }\href {\doibase 10.1103/PhysRevB.104.125113} {\bibfield
  {journal} {\bibinfo  {journal} {Phys. Rev. B}\ }\textbf {\bibinfo {volume}
  {104}},\ \bibinfo {pages} {125113} (\bibinfo {year} {2021})}\BibitemShut
  {NoStop}%
\bibitem [{\citenamefont {Li}\ \emph {et~al.}(2020)\citenamefont {Li},
  \citenamefont {Lee}, \citenamefont {Mu},\ and\ \citenamefont
  {Gong}}]{Li2020}%
  \BibitemOpen
  \bibfield  {author} {\bibinfo {author} {\bibfnamefont {Linhu}\ \bibnamefont
  {Li}}, \bibinfo {author} {\bibfnamefont {Ching~Hua}\ \bibnamefont {Lee}},
  \bibinfo {author} {\bibfnamefont {Sen}\ \bibnamefont {Mu}}, \ and\ \bibinfo
  {author} {\bibfnamefont {Jiangbin}\ \bibnamefont {Gong}},\ }\bibfield
  {title} {\enquote {\bibinfo {title} {Critical non-hermitian skin effect},}\
  }\href {\doibase 10.1038/s41467-020-18917-4} {\bibfield  {journal} {\bibinfo
  {journal} {Nature Communications}\ }\textbf {\bibinfo {volume} {11}},\
  \bibinfo {pages} {5491} (\bibinfo {year} {2020})}\BibitemShut {NoStop}%
\bibitem [{\citenamefont {Li}\ \emph {et~al.}(2021)\citenamefont {Li},
  \citenamefont {Lee},\ and\ \citenamefont {Gong}}]{Li2021}%
  \BibitemOpen
  \bibfield  {author} {\bibinfo {author} {\bibfnamefont {Linhu}\ \bibnamefont
  {Li}}, \bibinfo {author} {\bibfnamefont {Ching~Hua}\ \bibnamefont {Lee}}, \
  and\ \bibinfo {author} {\bibfnamefont {Jiangbin}\ \bibnamefont {Gong}},\
  }\bibfield  {title} {\enquote {\bibinfo {title} {Impurity induced scale-free
  localization},}\ }\href {\doibase 10.1038/s42005-021-00547-x} {\bibfield
  {journal} {\bibinfo  {journal} {Communications Physics}\ }\textbf {\bibinfo
  {volume} {4}},\ \bibinfo {pages} {42} (\bibinfo {year} {2021})}\BibitemShut
  {NoStop}%
\bibitem [{\citenamefont {Yokomizo}\ and\ \citenamefont
  {Murakami}(2021)}]{Kazuki2021}%
  \BibitemOpen
  \bibfield  {author} {\bibinfo {author} {\bibfnamefont {Kazuki}\ \bibnamefont
  {Yokomizo}}\ and\ \bibinfo {author} {\bibfnamefont {Shuichi}\ \bibnamefont
  {Murakami}},\ }\bibfield  {title} {\enquote {\bibinfo {title} {Scaling rule
  for the critical non-hermitian skin effect},}\ }\href {\doibase
  10.1103/PhysRevB.104.165117} {\bibfield  {journal} {\bibinfo  {journal}
  {Phys. Rev. B}\ }\textbf {\bibinfo {volume} {104}},\ \bibinfo {pages}
  {165117} (\bibinfo {year} {2021})}\BibitemShut {NoStop}%
\bibitem [{\citenamefont {Guo}\ \emph {et~al.}(2023)\citenamefont {Guo},
  \citenamefont {Wang}, \citenamefont {Hu},\ and\ \citenamefont
  {Chen}}]{Guo2023scalefree}%
  \BibitemOpen
  \bibfield  {author} {\bibinfo {author} {\bibfnamefont {Cui-Xian}\
  \bibnamefont {Guo}}, \bibinfo {author} {\bibfnamefont {Xueliang}\
  \bibnamefont {Wang}}, \bibinfo {author} {\bibfnamefont {Haiping}\
  \bibnamefont {Hu}}, \ and\ \bibinfo {author} {\bibfnamefont {Shu}\
  \bibnamefont {Chen}},\ }\bibfield  {title} {\enquote {\bibinfo {title}
  {Accumulation of scale-free localized states induced by local
  non-hermiticity},}\ }\href {\doibase 10.1103/PhysRevB.107.134121} {\bibfield
  {journal} {\bibinfo  {journal} {Phys. Rev. B}\ }\textbf {\bibinfo {volume}
  {107}},\ \bibinfo {pages} {134121} (\bibinfo {year} {2023})}\BibitemShut
  {NoStop}%
\bibitem [{dis()}]{distinction}%
  \BibitemOpen
  \href@noop {} {}\bibinfo {note} {Recently, we noticed that scale-free
  localization induced by a local loss is also discussed in
  Ref.~\cite{Guo2023scalefree} for a closed chain, which is consistent with our
  model under periodic boundary condition in Supplemental
  Material~\cite{supplemental}. However, the interplay between PT symmetry
  breaking and boundary conditions, and the subsequent criterion, are not
  studied in Ref.~\cite{Guo2023scalefree}. Moreover, the exsitence of a group
  of exponentially localized states induced by a single impurity is not found
  in Ref.~\cite{Guo2023scalefree}.}\BibitemShut {Stop}%
\bibitem [{\citenamefont {Yokomizo}\ and\ \citenamefont
  {Murakami}(2019)}]{Yokomizo2019}%
  \BibitemOpen
  \bibfield  {author} {\bibinfo {author} {\bibfnamefont {Kazuki}\ \bibnamefont
  {Yokomizo}}\ and\ \bibinfo {author} {\bibfnamefont {Shuichi}\ \bibnamefont
  {Murakami}},\ }\bibfield  {title} {\enquote {\bibinfo {title} {Non-bloch band
  theory of non-hermitian systems},}\ }\href {\doibase
  10.1103/PhysRevLett.123.066404} {\bibfield  {journal} {\bibinfo  {journal}
  {Phys. Rev. Lett.}\ }\textbf {\bibinfo {volume} {123}},\ \bibinfo {pages}
  {066404} (\bibinfo {year} {2019})}\BibitemShut {NoStop}%
\bibitem [{sup()}]{supplemental}%
  \BibitemOpen
  \href@noop {} {}\bibinfo {howpublished} {See Supplemental
  Material.}\BibitemShut {Stop}%
\bibitem [{\citenamefont {Makris}\ \emph {et~al.}(2008)\citenamefont {Makris},
  \citenamefont {El-Ganainy}, \citenamefont {Christodoulides},\ and\
  \citenamefont {Musslimani}}]{makris2008beam}%
  \BibitemOpen
  \bibfield  {author} {\bibinfo {author} {\bibfnamefont {K.~G.}\ \bibnamefont
  {Makris}}, \bibinfo {author} {\bibfnamefont {R.}~\bibnamefont {El-Ganainy}},
  \bibinfo {author} {\bibfnamefont {D.~N.}\ \bibnamefont {Christodoulides}}, \
  and\ \bibinfo {author} {\bibfnamefont {Z.~H.}\ \bibnamefont {Musslimani}},\
  }\bibfield  {title} {\enquote {\bibinfo {title} {Beam dynamics in
  $\mathcal{P}\mathcal{T}$ symmetric optical lattices},}\ }\href {\doibase
  10.1103/PhysRevLett.100.103904} {\bibfield  {journal} {\bibinfo  {journal}
  {Phys. Rev. Lett.}\ }\textbf {\bibinfo {volume} {100}},\ \bibinfo {pages}
  {103904} (\bibinfo {year} {2008})}\BibitemShut {NoStop}%
\bibitem [{\citenamefont {El-Ganainy}\ \emph {et~al.}(2007)\citenamefont
  {El-Ganainy}, \citenamefont {Makris}, \citenamefont {Christodoulides},\ and\
  \citenamefont {Musslimani}}]{El-Ganainy2007}%
  \BibitemOpen
  \bibfield  {author} {\bibinfo {author} {\bibfnamefont {R.}~\bibnamefont
  {El-Ganainy}}, \bibinfo {author} {\bibfnamefont {K.~G.}\ \bibnamefont
  {Makris}}, \bibinfo {author} {\bibfnamefont {D.~N.}\ \bibnamefont
  {Christodoulides}}, \ and\ \bibinfo {author} {\bibfnamefont {Ziad~H.}\
  \bibnamefont {Musslimani}},\ }\bibfield  {title} {\enquote {\bibinfo {title}
  {Theory of coupled optical pt-symmetric structures},}\ }\href {\doibase
  10.1364/OL.32.002632} {\bibfield  {journal} {\bibinfo  {journal} {Opt.
  Lett.}\ }\textbf {\bibinfo {volume} {32}},\ \bibinfo {pages} {2632--2634}
  (\bibinfo {year} {2007})}\BibitemShut {NoStop}%
\bibitem [{\citenamefont {R{\"u}ter}\ \emph {et~al.}(2010)\citenamefont
  {R{\"u}ter}, \citenamefont {Makris}, \citenamefont {El-Ganainy},
  \citenamefont {Christodoulides}, \citenamefont {Segev},\ and\ \citenamefont
  {Kip}}]{Ruter2010}%
  \BibitemOpen
  \bibfield  {author} {\bibinfo {author} {\bibfnamefont {Christian~E.}\
  \bibnamefont {R{\"u}ter}}, \bibinfo {author} {\bibfnamefont
  {Konstantinos~G.}\ \bibnamefont {Makris}}, \bibinfo {author} {\bibfnamefont
  {Ramy}\ \bibnamefont {El-Ganainy}}, \bibinfo {author} {\bibfnamefont
  {Demetrios~N.}\ \bibnamefont {Christodoulides}}, \bibinfo {author}
  {\bibfnamefont {Mordechai}\ \bibnamefont {Segev}}, \ and\ \bibinfo {author}
  {\bibfnamefont {Detlef}\ \bibnamefont {Kip}},\ }\bibfield  {title} {\enquote
  {\bibinfo {title} {Observation of parity--time symmetry in optics},}\ }\href
  {\doibase 10.1038/nphys1515} {\bibfield  {journal} {\bibinfo  {journal}
  {Nature Physics}\ }\textbf {\bibinfo {volume} {6}},\ \bibinfo {pages}
  {192--195} (\bibinfo {year} {2010})}\BibitemShut {NoStop}%
\bibitem [{\citenamefont {Guo}\ \emph {et~al.}(2009)\citenamefont {Guo},
  \citenamefont {Salamo}, \citenamefont {Duchesne}, \citenamefont {Morandotti},
  \citenamefont {Volatier-Ravat}, \citenamefont {Aimez}, \citenamefont
  {Siviloglou},\ and\ \citenamefont {Christodoulides}}]{guo2009complex}%
  \BibitemOpen
  \bibfield  {author} {\bibinfo {author} {\bibfnamefont {A.}~\bibnamefont
  {Guo}}, \bibinfo {author} {\bibfnamefont {G.~J.}\ \bibnamefont {Salamo}},
  \bibinfo {author} {\bibfnamefont {D.}~\bibnamefont {Duchesne}}, \bibinfo
  {author} {\bibfnamefont {R.}~\bibnamefont {Morandotti}}, \bibinfo {author}
  {\bibfnamefont {M.}~\bibnamefont {Volatier-Ravat}}, \bibinfo {author}
  {\bibfnamefont {V.}~\bibnamefont {Aimez}}, \bibinfo {author} {\bibfnamefont
  {G.~A.}\ \bibnamefont {Siviloglou}}, \ and\ \bibinfo {author} {\bibfnamefont
  {D.~N.}\ \bibnamefont {Christodoulides}},\ }\bibfield  {title} {\enquote
  {\bibinfo {title} {Observation of $\mathcal{P}\mathcal{T}$-symmetry breaking
  in complex optical potentials},}\ }\href {\doibase
  10.1103/PhysRevLett.103.093902} {\bibfield  {journal} {\bibinfo  {journal}
  {Phys. Rev. Lett.}\ }\textbf {\bibinfo {volume} {103}},\ \bibinfo {pages}
  {093902} (\bibinfo {year} {2009})}\BibitemShut {NoStop}%
\bibitem [{\citenamefont {Zhen}\ \emph {et~al.}(2015)\citenamefont {Zhen},
  \citenamefont {Hsu}, \citenamefont {Igarashi}, \citenamefont {Lu},
  \citenamefont {Kaminer}, \citenamefont {Pick}, \citenamefont {Chua},
  \citenamefont {Joannopoulos},\ and\ \citenamefont
  {Solja{\v{c}}i{\'{c}}}}]{Zhen2015}%
  \BibitemOpen
  \bibfield  {author} {\bibinfo {author} {\bibfnamefont {Bo}~\bibnamefont
  {Zhen}}, \bibinfo {author} {\bibfnamefont {Chia~Wei}\ \bibnamefont {Hsu}},
  \bibinfo {author} {\bibfnamefont {Yuichi}\ \bibnamefont {Igarashi}}, \bibinfo
  {author} {\bibfnamefont {Ling}\ \bibnamefont {Lu}}, \bibinfo {author}
  {\bibfnamefont {Ido}\ \bibnamefont {Kaminer}}, \bibinfo {author}
  {\bibfnamefont {Adi}\ \bibnamefont {Pick}}, \bibinfo {author} {\bibfnamefont
  {Song-Liang}\ \bibnamefont {Chua}}, \bibinfo {author} {\bibfnamefont
  {John~D.}\ \bibnamefont {Joannopoulos}}, \ and\ \bibinfo {author}
  {\bibfnamefont {Marin}\ \bibnamefont {Solja{\v{c}}i{\'{c}}}},\ }\bibfield
  {title} {\enquote {\bibinfo {title} {Spawning rings of exceptional points out
  of dirac cones},}\ }\href {\doibase 10.1038/nature14889} {\bibfield
  {journal} {\bibinfo  {journal} {Nature}\ }\textbf {\bibinfo {volume} {525}},\
  \bibinfo {pages} {354--358} (\bibinfo {year} {2015})}\BibitemShut {NoStop}%
\bibitem [{\citenamefont {Ding}\ \emph {et~al.}(2015)\citenamefont {Ding},
  \citenamefont {Zhang},\ and\ \citenamefont {Chan}}]{Ding2015}%
  \BibitemOpen
  \bibfield  {author} {\bibinfo {author} {\bibfnamefont {Kun}\ \bibnamefont
  {Ding}}, \bibinfo {author} {\bibfnamefont {Z.~Q.}\ \bibnamefont {Zhang}}, \
  and\ \bibinfo {author} {\bibfnamefont {C.~T.}\ \bibnamefont {Chan}},\
  }\bibfield  {title} {\enquote {\bibinfo {title} {Coalescence of exceptional
  points and phase diagrams for one-dimensional
  $\mathcal{P}\mathcal{T}$-symmetric photonic crystals},}\ }\href {\doibase
  10.1103/PhysRevB.92.235310} {\bibfield  {journal} {\bibinfo  {journal} {Phys.
  Rev. B}\ }\textbf {\bibinfo {volume} {92}},\ \bibinfo {pages} {235310}
  (\bibinfo {year} {2015})}\BibitemShut {NoStop}%
\bibitem [{\citenamefont {Cerjan}\ \emph {et~al.}(2016)\citenamefont {Cerjan},
  \citenamefont {Raman},\ and\ \citenamefont {Fan}}]{Cerjan2016}%
  \BibitemOpen
  \bibfield  {author} {\bibinfo {author} {\bibfnamefont {Alexander}\
  \bibnamefont {Cerjan}}, \bibinfo {author} {\bibfnamefont {Aaswath}\
  \bibnamefont {Raman}}, \ and\ \bibinfo {author} {\bibfnamefont {Shanhui}\
  \bibnamefont {Fan}},\ }\bibfield  {title} {\enquote {\bibinfo {title}
  {Exceptional contours and band structure design in parity-time symmetric
  photonic crystals},}\ }\href {\doibase 10.1103/PhysRevLett.116.203902}
  {\bibfield  {journal} {\bibinfo  {journal} {Phys. Rev. Lett.}\ }\textbf
  {\bibinfo {volume} {116}},\ \bibinfo {pages} {203902} (\bibinfo {year}
  {2016})}\BibitemShut {NoStop}%
\bibitem [{\citenamefont {Carusotto}\ \emph {et~al.}(2009)\citenamefont
  {Carusotto}, \citenamefont {Gerace}, \citenamefont {Tureci}, \citenamefont
  {De~Liberato}, \citenamefont {Ciuti},\ and\ \citenamefont
  {Imamo\ifmmode~\check{g}\else \v{g}\fi{}lu}}]{Carusotto2009}%
  \BibitemOpen
  \bibfield  {author} {\bibinfo {author} {\bibfnamefont {I.}~\bibnamefont
  {Carusotto}}, \bibinfo {author} {\bibfnamefont {D.}~\bibnamefont {Gerace}},
  \bibinfo {author} {\bibfnamefont {H.~E.}\ \bibnamefont {Tureci}}, \bibinfo
  {author} {\bibfnamefont {S.}~\bibnamefont {De~Liberato}}, \bibinfo {author}
  {\bibfnamefont {C.}~\bibnamefont {Ciuti}}, \ and\ \bibinfo {author}
  {\bibfnamefont {A.}~\bibnamefont {Imamo\ifmmode~\check{g}\else
  \v{g}\fi{}lu}},\ }\bibfield  {title} {\enquote {\bibinfo {title} {Fermionized
  photons in an array of driven dissipative nonlinear cavities},}\ }\href
  {\doibase 10.1103/PhysRevLett.103.033601} {\bibfield  {journal} {\bibinfo
  {journal} {Phys. Rev. Lett.}\ }\textbf {\bibinfo {volume} {103}},\ \bibinfo
  {pages} {033601} (\bibinfo {year} {2009})}\BibitemShut {NoStop}%
\bibitem [{\citenamefont {Umucal\ifmmode \imath \else~\i \fi{}lar}\ and\
  \citenamefont {Carusotto}(2012)}]{Umucal2012}%
  \BibitemOpen
  \bibfield  {author} {\bibinfo {author} {\bibfnamefont {R.~O.}\ \bibnamefont
  {Umucal\ifmmode \imath \else~\i \fi{}lar}}\ and\ \bibinfo {author}
  {\bibfnamefont {I.}~\bibnamefont {Carusotto}},\ }\bibfield  {title} {\enquote
  {\bibinfo {title} {Fractional quantum hall states of photons in an array of
  dissipative coupled cavities},}\ }\href {\doibase
  10.1103/PhysRevLett.108.206809} {\bibfield  {journal} {\bibinfo  {journal}
  {Phys. Rev. Lett.}\ }\textbf {\bibinfo {volume} {108}},\ \bibinfo {pages}
  {206809} (\bibinfo {year} {2012})}\BibitemShut {NoStop}%
\bibitem [{\citenamefont {Bardyn}\ and\ \citenamefont {\ifmmode \dot{I}\else
  \.{I}\fi{}mamo\ifmmode~\check{g}\else \v{g}\fi{}lu}(2012)}]{Bardyn2012}%
  \BibitemOpen
  \bibfield  {author} {\bibinfo {author} {\bibfnamefont {C.-E.}\ \bibnamefont
  {Bardyn}}\ and\ \bibinfo {author} {\bibfnamefont {A.}~\bibnamefont {\ifmmode
  \dot{I}\else \.{I}\fi{}mamo\ifmmode~\check{g}\else \v{g}\fi{}lu}},\
  }\bibfield  {title} {\enquote {\bibinfo {title} {Majorana-like modes of light
  in a one-dimensional array of nonlinear cavities},}\ }\href {\doibase
  10.1103/PhysRevLett.109.253606} {\bibfield  {journal} {\bibinfo  {journal}
  {Phys. Rev. Lett.}\ }\textbf {\bibinfo {volume} {109}},\ \bibinfo {pages}
  {253606} (\bibinfo {year} {2012})}\BibitemShut {NoStop}%
\bibitem [{\citenamefont {Dartiailh}\ \emph {et~al.}(2017)\citenamefont
  {Dartiailh}, \citenamefont {Kontos}, \citenamefont
  {Dou\ifmmode~\mbox{\c{c}}\else \c{c}\fi{}ot},\ and\ \citenamefont
  {Cottet}}]{Dartiailh2017}%
  \BibitemOpen
  \bibfield  {author} {\bibinfo {author} {\bibfnamefont {Matthieu~C.}\
  \bibnamefont {Dartiailh}}, \bibinfo {author} {\bibfnamefont {Takis}\
  \bibnamefont {Kontos}}, \bibinfo {author} {\bibfnamefont {Benoit}\
  \bibnamefont {Dou\ifmmode~\mbox{\c{c}}\else \c{c}\fi{}ot}}, \ and\ \bibinfo
  {author} {\bibfnamefont {Audrey}\ \bibnamefont {Cottet}},\ }\bibfield
  {title} {\enquote {\bibinfo {title} {Direct cavity detection of majorana
  pairs},}\ }\href {\doibase 10.1103/PhysRevLett.118.126803} {\bibfield
  {journal} {\bibinfo  {journal} {Phys. Rev. Lett.}\ }\textbf {\bibinfo
  {volume} {118}},\ \bibinfo {pages} {126803} (\bibinfo {year}
  {2017})}\BibitemShut {NoStop}%
\bibitem [{\citenamefont {Deuar}\ \emph {et~al.}(2021)\citenamefont {Deuar},
  \citenamefont {Ferrier}, \citenamefont {Matuszewski}, \citenamefont {Orso},\
  and\ \citenamefont {Szyma\ifmmode~\acute{n}\else \'{n}\fi{}ska}}]{Deuar2021}%
  \BibitemOpen
  \bibfield  {author} {\bibinfo {author} {\bibfnamefont {Piotr}\ \bibnamefont
  {Deuar}}, \bibinfo {author} {\bibfnamefont {Alex}\ \bibnamefont {Ferrier}},
  \bibinfo {author} {\bibfnamefont {Micha\l{}}\ \bibnamefont {Matuszewski}},
  \bibinfo {author} {\bibfnamefont {Giuliano}\ \bibnamefont {Orso}}, \ and\
  \bibinfo {author} {\bibfnamefont {Marzena~H.}\ \bibnamefont
  {Szyma\ifmmode~\acute{n}\else \'{n}\fi{}ska}},\ }\bibfield  {title} {\enquote
  {\bibinfo {title} {Fully quantum scalable description of driven-dissipative
  lattice models},}\ }\href {\doibase 10.1103/PRXQuantum.2.010319} {\bibfield
  {journal} {\bibinfo  {journal} {PRX Quantum}\ }\textbf {\bibinfo {volume}
  {2}},\ \bibinfo {pages} {010319} (\bibinfo {year} {2021})}\BibitemShut
  {NoStop}%
\bibitem [{\citenamefont {Alcaraz}\ \emph {et~al.}(1987)\citenamefont
  {Alcaraz}, \citenamefont {Barber}, \citenamefont {Batchelor}, \citenamefont
  {Baxter},\ and\ \citenamefont {Quispel}}]{FCAlcaraz1987}%
  \BibitemOpen
  \bibfield  {author} {\bibinfo {author} {\bibfnamefont {F~C}\ \bibnamefont
  {Alcaraz}}, \bibinfo {author} {\bibfnamefont {M~N}\ \bibnamefont {Barber}},
  \bibinfo {author} {\bibfnamefont {M~T}\ \bibnamefont {Batchelor}}, \bibinfo
  {author} {\bibfnamefont {R~J}\ \bibnamefont {Baxter}}, \ and\ \bibinfo
  {author} {\bibfnamefont {G~R~W}\ \bibnamefont {Quispel}},\ }\bibfield
  {title} {\enquote {\bibinfo {title} {Surface exponents of the quantum xxz,
  ashkin-teller and potts models},}\ }\href {\doibase
  10.1088/0305-4470/20/18/038} {\bibfield  {journal} {\bibinfo  {journal}
  {Journal of Physics A: Mathematical and General}\ }\textbf {\bibinfo {volume}
  {20}},\ \bibinfo {pages} {6397} (\bibinfo {year} {1987})}\BibitemShut
  {NoStop}%
\bibitem [{\citenamefont {Wang}\ \emph {et~al.}(2023)\citenamefont {Wang},
  \citenamefont {Li}, \citenamefont {Song},\ and\ \citenamefont
  {Wang}}]{Wang2023Scale}%
  \BibitemOpen
  \bibfield  {author} {\bibinfo {author} {\bibfnamefont {He-Ran}\ \bibnamefont
  {Wang}}, \bibinfo {author} {\bibfnamefont {Bo}~\bibnamefont {Li}}, \bibinfo
  {author} {\bibfnamefont {Fei}\ \bibnamefont {Song}}, \ and\ \bibinfo {author}
  {\bibfnamefont {Zhong}\ \bibnamefont {Wang}},\ }\bibfield  {title} {\enquote
  {\bibinfo {title} {Scale-free non-hermitian skin effect in a
  boundary-dissipated spin chain},}\ }\href
  {https://arxiv.org/abs/2301.11896v3} {\bibfield  {journal} {\bibinfo
  {journal} {arXiv:2301.11896v3}\ } (\bibinfo {year} {2023})}\BibitemShut
  {NoStop}%
\end{thebibliography}%


\begin{thebibliography}{2}%
\makeatletter
\providecommand \@ifxundefined [1]{%
 \@ifx{#1\undefined}
}%
\providecommand \@ifnum [1]{%
 \ifnum #1\expandafter \@firstoftwo
 \else \expandafter \@secondoftwo
 \fi
}%
\providecommand \@ifx [1]{%
 \ifx #1\expandafter \@firstoftwo
 \else \expandafter \@secondoftwo
 \fi
}%
\providecommand \natexlab [1]{#1}%
\providecommand \enquote  [1]{``#1''}%
\providecommand \bibnamefont  [1]{#1}%
\providecommand \bibfnamefont [1]{#1}%
\providecommand \citenamefont [1]{#1}%
\providecommand \href@noop [0]{\@secondoftwo}%
\providecommand \href [0]{\begingroup \@sanitize@url \@href}%
\providecommand \@href[1]{\@@startlink{#1}\@@href}%
\providecommand \@@href[1]{\endgroup#1\@@endlink}%
\providecommand \@sanitize@url [0]{\catcode `\\12\catcode `\$12\catcode
  `\&12\catcode `\#12\catcode `\^12\catcode `\_12\catcode `\%12\relax}%
\providecommand \@@startlink[1]{}%
\providecommand \@@endlink[0]{}%
\providecommand \url  [0]{\begingroup\@sanitize@url \@url }%
\providecommand \@url [1]{\endgroup\@href {#1}{\urlprefix }}%
\providecommand \urlprefix  [0]{URL }%
\providecommand \Eprint [0]{\href }%
\providecommand \doibase [0]{http://dx.doi.org/}%
\providecommand \selectlanguage [0]{\@gobble}%
\providecommand \bibinfo  [0]{\@secondoftwo}%
\providecommand \bibfield  [0]{\@secondoftwo}%
\providecommand \translation [1]{[#1]}%
\providecommand \BibitemOpen [0]{}%
\providecommand \bibitemStop [0]{}%
\providecommand \bibitemNoStop [0]{.\EOS\space}%
\providecommand \EOS [0]{\spacefactor3000\relax}%
\providecommand \BibitemShut  [1]{\csname bibitem#1\endcsname}%
\let\auto@bib@innerbib\@empty
\bibitem [{\citenamefont {Yao}\ and\ \citenamefont {Wang}(2018)}]{yao2018edge}%
  \BibitemOpen
  \bibfield  {author} {\bibinfo {author} {\bibfnamefont {Shunyu}\ \bibnamefont
  {Yao}}\ and\ \bibinfo {author} {\bibfnamefont {Zhong}\ \bibnamefont {Wang}},\
  }\bibfield  {title} {\enquote {\bibinfo {title} {Edge states and topological
  invariants of non-hermitian systems},}\ }\href {\doibase
  10.1103/PhysRevLett.121.086803} {\bibfield  {journal} {\bibinfo  {journal}
  {Phys. Rev. Lett.}\ }\textbf {\bibinfo {volume} {121}},\ \bibinfo {pages}
  {086803} (\bibinfo {year} {2018})}\BibitemShut {NoStop}%
\bibitem [{\citenamefont {Yokomizo}\ and\ \citenamefont
  {Murakami}(2019)}]{Yokomizo2019}%
  \BibitemOpen
  \bibfield  {author} {\bibinfo {author} {\bibfnamefont {Kazuki}\ \bibnamefont
  {Yokomizo}}\ and\ \bibinfo {author} {\bibfnamefont {Shuichi}\ \bibnamefont
  {Murakami}},\ }\bibfield  {title} {\enquote {\bibinfo {title} {Non-bloch band
  theory of non-hermitian systems},}\ }\href {\doibase
  10.1103/PhysRevLett.123.066404} {\bibfield  {journal} {\bibinfo  {journal}
  {Phys. Rev. Lett.}\ }\textbf {\bibinfo {volume} {123}},\ \bibinfo {pages}
  {066404} (\bibinfo {year} {2019})}\BibitemShut {NoStop}%
\end{thebibliography}%

\end{document}


\title{Supplemental Material for ``Scale-free localization and PT symmetry breaking from local non-Hermiticity"}

\author{Bo Li}
 \affiliation{ Institute for
Advanced Study, Tsinghua University, Beijing,  100084, China }
 \author{He-Ran Wang}
 \affiliation{ Institute for
Advanced Study, Tsinghua University, Beijing,  100084, China }
\author{Fei Song}
\affiliation{ Institute for
Advanced Study, Tsinghua University, Beijing,  100084, China }
\author{Zhong Wang} 
\affiliation{ Institute for
Advanced Study, Tsinghua University, Beijing,  100084, China }

\maketitle
\onecolumngrid

%
%
%
%
%
%
%
%
%

\section{Scale-free localization for a simple chain model}

\begin{figure}
\includegraphics[width=0.4\linewidth]{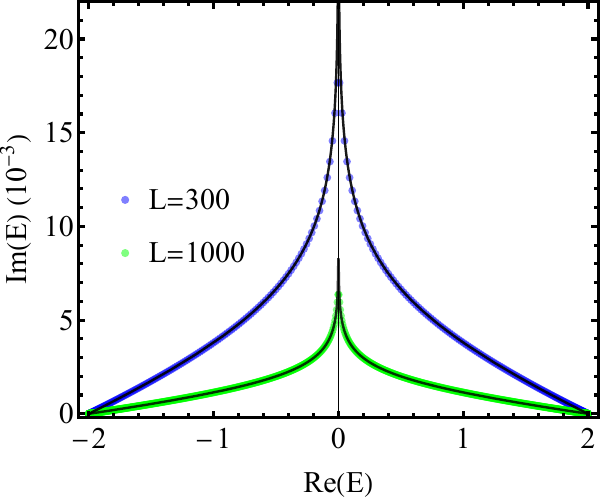}
 \caption{Eigenvalues (of model Eq.(1) in main text ) in complex plane reproduced by using the scale-free ansatz Eq.~\eqref{seq:ansatz}. Here, the dots represents eigenvalues obtained from exact diagonalization; the black lines are given by parametric plot according to Eq.~\eqref{seq:scaleconstant1} and \eqref{seq:paraeigenvalue}, where eigenvalues are reserved to the leading order of $L^{-1}$.  }\label{fig:scale_constant}
\end{figure}

For the chain model presented in Eq.(1) in the main text,  the scale-free localization of eigenstates implies that the boundary condition can be satisfied, up to $O(L^{-1})$, by the solution of form
\begin{eqnarray}\label{seq:ansatz}
    \beta=e^{c/L+i\theta},
\end{eqnarray}
where $c$ is a real parameter.
Substituting this ansatz into the boundary condition in the main text, i.e.,
\begin{eqnarray}\label{eq:boundarycondition}
\det\left(
\begin{array}{cc}
ig\beta-t & ig\beta^{-1}-t\\
t\beta^{L+1} &t\beta^{-(L+1)}
\end{array}
\right)=0,
\end{eqnarray}
the parameter $c$ can be solved by keeping the leading order of the equation, which gives
\begin{eqnarray}\label{seq:scaleconstant}
    e^{2c}=\frac{ig e^{i\theta}-t}{ig e^{-i\theta}-t}e^{-i2(L+1)\theta}.
\end{eqnarray}
By taking absolute value on both side, the value of $c$ is given by
\begin{eqnarray}\label{seq:scaleconstant1}
    c=\frac{1}{2}\ln{\frac{\sqrt{(t^2+g^2\cos(2\theta))^2+g^4\sin^2(2\theta)}}{g^2+t^2-2gt\sin\theta}}.
\end{eqnarray}
For a lattice model, the value of $\theta$ takes discrete values to guarantee the r.h.s of Eq.~\eqref{seq:scaleconstant} to be real. A convenient way to check the expression Eq.~\eqref{seq:scaleconstant1} is to reproduce the eigenvalues, i.e., $E=t(\beta+\beta^{-1})$. To the leading order of $L^{-1}$, the eigenvalues can be decomposed into real and imaginary part 
\begin{eqnarray}\label{seq:paraeigenvalue}
    (\text{Re}(E),\text{Im}(E))=(2t\cos\theta, 2tc \sin\theta/L),
\end{eqnarray}
which is plotted parametrically in Fig.~\ref{fig:scale_constant}. The parametric plot agrees well with the exact diagonalization result, confirming the expression of scale constant $c$.

\section{Accumulation of bound states due to local non-Hermitian impurity}
\begin{figure}
    \centering
\begin{tabular}{cc}
\includegraphics[width=0.8\linewidth]{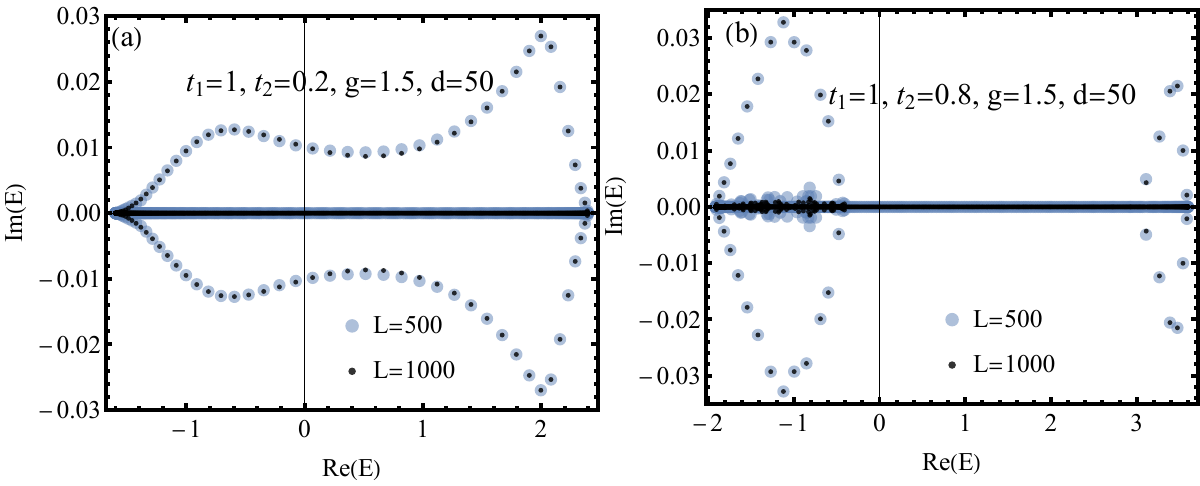}\\
\includegraphics[width=0.8\linewidth]{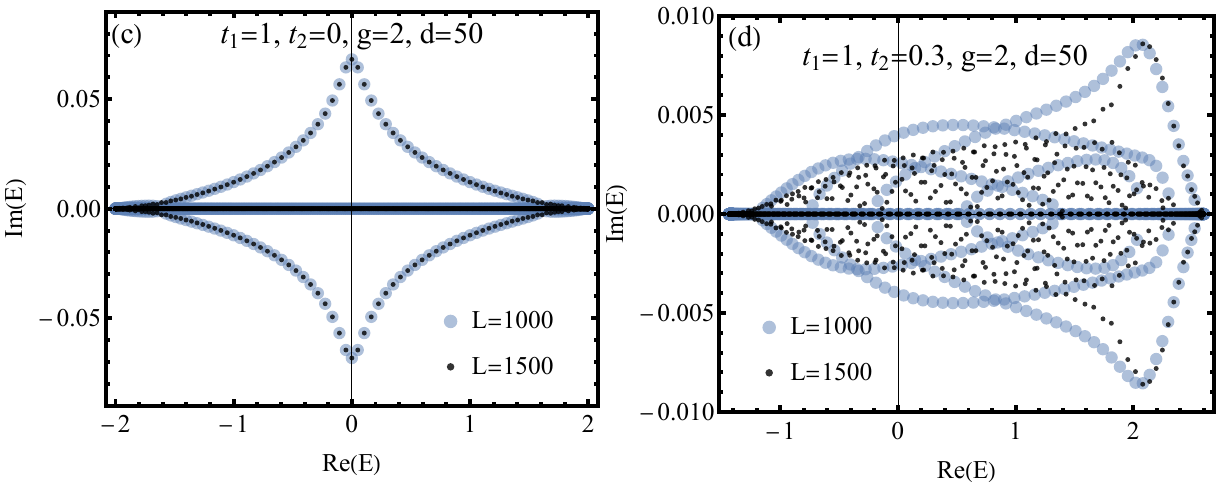}
\end{tabular}
\caption{(a),(b) Spectrum for the OBC case with PT symmetric impurities respectively lying at sites with distance $d$ to two ends. (c),(d) Spectrum for PBC case with PT symmetric impurities siting $2d$ away from each other.}
\label{fig:twoimpurities}
\end{figure}
In this section, we give more details of the model discussed in the main text, and then extend the discussion to general situation. To be clear, we first explicitly write down the model
\begin{eqnarray}
H=\sum_{j=1}^{L-1}t(|j\rangle\langle j+1|+h.c.)+ig|d\rangle\langle d|.
\end{eqnarray}
The eigenvalues here is characterized by $E=t(\beta+\beta^{-1})$, while the eigenstate has a sudden change at the impurity, $|\psi\rangle=\sum_{1\leq j<d}\psi_j^{(1)}|j\rangle+\sum_{d\leq j\leq L}\psi_j^{(2)}|j\rangle$, with $\psi^{(\nu)}_j=a_\nu\beta^{j}+b_\nu\beta^{-j}$ ($\nu=1,2$). The coefficient $a_\nu, b_\nu$ can be determined by boundary condition and bulk equation across the impurity. The eigen equation of the system is $t\psi_{j-1}+t\psi_{j+1}=E\psi_j$ except for sites near the impurity and boundary. To accommodate the eigen equation at boundaries, the wavefunction has to be vanish outside of the chain  
\begin{eqnarray}
&&\psi_0=a_1+b_1=0,\label{eq:bound_cond1} \\
&&\psi_{L+1}=a_2\beta^{L+1}+b_2\beta^{-{L+1}}=0.\label{eq:bound_cond2}
\end{eqnarray}
Near the impurity, we need to take care of the imaginary potential and the sudden change of combination coefficients, which yields the following constraints
\begin{eqnarray}
t\psi^{(1)}_{d-1}+ig\psi_d^{(2)}+t\psi_{d+1}^{(2)}=E(\beta)\psi_d^{(2)},\label{eq:bound_cond3}\\
t\psi_{d-2}^{(1)}+t\psi_{d}^{(2)}=E(\beta)\psi_{d-1}^{(1)}\label{eq:bound_cond4}.
\end{eqnarray}
If only the bound states near the impurity are concerned, namely decaying from site $d$ to $L$, we can assume $|\beta|<1$. In the thermodynamic limit $L\rightarrow \infty$, Eq.~\eqref{eq:bound_cond2} demands $a_2=0$. Now, we are left with three equations and three free parameters, and the information of system size $L$ is eliminated. By applying the ansatz to the rest equations, it is straightforward to reach the following condition,
\begin{eqnarray}
ig\beta^{2d+1}-t\beta^2-ig\beta+t=0.
\end{eqnarray}\\
\textbf{More general case}. The core of above discussion is that for an impurity leaving one boundary with distance $d$, we only focus on bound states that exponentially localize to the impurity, which makes these states only sensitive to the information between the impurity and the closer boundary. As a result, the equations of these bound states of concern have highest order linear in $d$, and thus the number of satisfactory solutions is $\mathcal O(d)$. These bound states exist in the thermodynamic limit and its number can be controlled by the distance $d$.\\
\\
The same logic works as well  when adding non-Hermitian impurities at two sides of an open chain with corresponding distance $d_1, d_2$ to their closer boundary. In the thermodynamic limit, i.e., two impurities infinitely far from each other, the bound states accumulate at different impurites decouple with each other. Therefore, the problem for each impurity is reduced to the case before, and the number of bound states is controlled by $d_1, d_2$, respectively. Moreover, if two impurities with interval $d$ are added on a closed chain or deep inside of an open chain (infinitely far from both boundaries when $L\rightarrow\infty$), the equations to solve only involve more coefficients to account the two impurities, but the highest order of equation is still $\mathcal O(d)$. Hence, the situation is again analogous to what we have solved, and exponentially localized states exist in the thermodynamic limit.\\
\\
The scenario becomes complicated when longer range hopping is included, however the underlying origin of bound state is the same as before. One key observation is that there should be a set of complex  eigenvalues whose amount is determined by the distance $d$  (or $d_1, d_2$) even if $L\rightarrow\infty$. This can be seen as the following: the eigenvalues in general are continuous functions of $\{t_n\}$, the hopping parameters; a model with further hopping range, i.e., $t_{n>1}\neq 0$, can be always continuously deformed to the simplest model with only nearest neighbor hopping ($t_{n>1}=0$), so do their eigenvalues; as we know there is a set of complex eigenvalues for the $t_{n>1}=0$ case, so models with $t_{n>1}\neq 0$ should in general has similar amount of complex eigenvalues, otherwise the imaginary part of complex eigenvalues will suddenly jump from nonzero to zero.  Already knowing existence of complex eigenvalues $\{E_\nu\}$, its corresponding characteristic equation $\sum_{n}t_n\beta^{n}+t^\ast_n\beta^{-n}=E_\nu$ in general gives solution not living on Brillioun zone, i.e., $|\beta|\neq 1$ and $\beta\in\mathbbm{C}$, which corresponds bound state.\\
\\
\textbf{More examples}. We give examples with second nearest neighbor hopping and PT symmetric impurities under both OBC and PBC. The Hamiltonians are respectively  given by
\begin{eqnarray}
&&H_{\text{OBC}}=\sum_{j=1}^{L-1} t_1(|j\rangle\langle j+1|+h.c.)+\sum_{j=1}^{L-2} t_2(|j\rangle\langle j+2|+h.c.)+ ig(|d\rangle\langle d|-|L-d\rangle\langle L-d|),\\
&&H_{\text{PBC}}=\sum_{j=1}^{L} (t_1|j\rangle\langle j+1|+ t_2|j\rangle\langle j+2|+h.c.)+ ig(|d\rangle\langle d|-|L-d+1\rangle\langle L-d+1|),
\end{eqnarray}
where $|j+L\rangle=|j\rangle$ is implies under PBC. Actually, the gain and loss with equal strength can be placed at any two sites under PBC, and the choice above is made for convenience. The spectrum with certain parameters for two cases are plotted in Fig.~\ref{fig:twoimpurities}. From (a), (b), we see that when impurities are not far from boundaries, PT breaking takes place for bound states ($\text{Im}(E)$ is independent of $L$) without restricted by the criterion in the main text, (i.e., appear even when $t_2<t_1/4$), while the scale-free modes ($\text{Im}(E)$ decreases with increasing $L$) still respect the criterion (only shows up when $t_2>t_1/4$). In the PBC case shown in (c),(d), a collection of bound states and the scale-free modes present at the same time. These two examples illustrate that the appearance of a group of bound states under appropriate arrangement of non-Hermitian impurities is a generic phenomenon, and it is not demanding on the details of the original system.

\section{Eigenstates for Hermitian systems perturbed by a local non-Hermitian term}

In this section, we analyze the feature of eigenstates for a system perturbed by a local non-Hermitian term. For concreteness, we assume that the magnitude of perturbation is comparable with the characteristic energy scale of original system.  First, we will show that the eigenstate is extended when eigenvalues are real.
Then, we concentrate the general case for a PT symmetric system to justify the existence of scale-free localization in the PT-broken phase.
\\

\subsection{Eigenstates for real eigenvalues}

In this part, we show that for a Hermitian system perturbed by non-Hermitian boundary (or a local non-Hermitian term), when its eigenvalues are real, the eigenstates for  continuous spectrum are always uniformly extended. 
\\

\textbf{OBC case}. We first study the OBC case. Suppose the Hamiltonian is generally given by
\begin{eqnarray}
H^{(o)}=\sum_{n=1}^M h^{(o)}_n+V,
\end{eqnarray}
where the first term represents hopping terms with different hopping range $n$ and $M$ refers to the maximum hopping range, the second term is a local non-Hermitian potential. The specific form of two terms are given below,
\begin{eqnarray}
h^{(o)}_n&&=\sum_{i=1}^{L-n}t_n|i\rangle\langle i+n|+ t_n^\ast |i+n\rangle\langle i|,\nonumber\\
V&&=\sum_{i\in\partial\Omega} v_i|i\rangle\langle i|+\sum_{i,j\in\partial\Omega}v_{ij}|i\rangle\langle j|,
\end{eqnarray}
where $t_n$ is a complex interaction between sites with distance $n$, $V^\dagger\neq V$, and $\partial\Omega$ stands for sites near the boundary (assuming the size of $\partial\Omega$ is smaller than the maximum hopping range $M$). The eigenstate of such system can be usually constructed based on the state $|\beta\rangle=\sum_j \beta^j|j\rangle$, where $\beta$ takes complex values. For our purpose, we explore the possibility that all satisfactory solutions ($\beta$) have non-unitary modulus. The two terms in Hamiltonian act on $|\beta\rangle$ respectively give
\begin{eqnarray}\label{eq:findeigenstate}
&&h^{(o)}_n|\beta\rangle=(t_n\beta^n+t_n^\ast \beta^{-n})|\beta\rangle-t_n\beta^L\sum_{l=1}^n \beta^l|L-(n-l)\rangle-t_n^\ast \beta^{-n}\sum_{l=1}^n\beta^l|l\rangle,\nonumber\\
&&V|\beta\rangle=\sum_{i\in\partial\Omega}v_i\beta^i|i\rangle+\sum_{i,j\in\partial\Omega}v_{ij}\beta^j|i\rangle.
\end{eqnarray}
The  continuous spectrum is generally given by
\begin{eqnarray}\label{eq:generalenergy}
E=\sum_{n=1}^M t_n\beta^n+t_{n}^\ast \beta^{-n}.
\end{eqnarray}
Here, we can see that the solutions form pairs $(\beta,1/\beta^\ast)$ for real energy. If solutions are ordered as $|\beta_1|\leq |\beta_2|\leq \cdots\leq |\beta_{2M}|$, it must have
$\beta_j=1/\beta_{2M-j+1}^\ast$ for $1\leq j\leq M$. In general, eigenstates can be constructed by linearly combining  states corresponding different solution $\beta_j$ as 
\begin{eqnarray}\label{eq:wavefunction}
|\psi\rangle=\sum_{i=1}^{2M}c_j |\beta_j\rangle.
\end{eqnarray}
Acting the Hamiltonian $H^{(o)}$ on $|\psi\rangle$ generates
\begin{eqnarray}\label{eq:eigenequation}
H^{(o)}|\psi\rangle= E|\psi\rangle +B^{(o)}
\end{eqnarray}
with 
\begin{eqnarray}
B^{(o)}=-\sum_{i=1}^M\sum_{n=i}^M\sum_{l=1}^{2M}c_l[t_n \beta_l^{L+i}|L-(n-i)\rangle+t_n^\ast \beta_l^{i-n}|i\rangle]+\sum_{l=1}^{2M}\sum_{i\in\partial\Omega}c_lv_i\beta_l^i|i\rangle+\sum_{l=1}^{2M}\sum_{i,j\in\partial\Omega}c_lv_{ij}\beta_l^j|i\rangle.
\end{eqnarray}
To satisfy the eigen-equation, the tail $B^{(o)}$ in Eq.~\eqref{eq:eigenequation} has to vanish, i.e., $B^{(o)}=0$, which can be realized by choosing appropriate combination  coefficients $c_l$. The corresponding condition is generally given by $F^{(o)}(\{\beta_i\})\cdot(c_1,c_2,\cdots,c_{2M})^T=0$ with
\begin{eqnarray}
F^{(o)}(\{\beta_i\})=\left(
\begin{array}{cccccc}
    f_1(\beta_1)&\cdots &f_1(\beta_{2M})\\ \vdots&\vdots&\vdots\\
    f_M(\beta_1)&\cdots &f_M(\beta_{2M})\\
    \beta_1^Lg_1(\beta_1)&\cdots &\beta_{2M}^Lg_1(\beta_{2M})\\ \vdots&\vdots&\vdots\\
    \beta_1^Lg_M(\beta_1)&\cdots &\beta_{2M}^Lg_M(\beta_{2M})
\end{array}
\right),
\end{eqnarray}
where $f_i(\beta_j)$, $g_i(\beta_j)$ are polynomials of $\beta_j$ with highest order less than the hopping range $M$.  The condition is satisfied if
\begin{eqnarray}
\det F^{(o)}(\{\beta_i\})=0,
\end{eqnarray}
which is the general boundary condition. The boundary condition demands the vanishing of a polynomial which has the leading and sub-leading term $(\beta_{2M}\beta_{2M-1}\cdots \beta_{M+2} \beta_{M+1})^L$ and $(\beta_{2M}\beta_{2M-1}\cdots \beta_{M+2} \beta_{M})^L$. For a large finte system $(L\gg 1)$, similar to the non-Bloch band theory~\cite{yao2018edge, Yokomizo2019}, it requires $|\beta_{M+1}|^L\sim |\beta_M|^L$ (not necessary to be exactly equal) to make the two terms cancel out. Moreover, the real-valued eigenvalues enforce that
\begin{eqnarray}
\beta_M=1/\beta_{M+1}^\ast.
\end{eqnarray}
Therefore, the boundary condition and real-valued continuous spectrum require
\begin{eqnarray}
|\beta_{M}|^L\sim|1/\beta^\ast_M|^L
\end{eqnarray}
which obviously demands that $|\beta_M|^L\sim 1$.  This implies that $|\beta_M|$ can be given by a Taylor expansion $|\beta_M|=1+\sum_{n=1}^\infty a_n/L^n$. Taking $\beta_M=|\beta_M|e^{i\theta}$, the eigenvalues can be obtained by plugging $\beta_M$ into Eq.~\eqref{eq:generalenergy}. And the real-valued spectrum requires vanishing imaginary part of eigenvalues
\begin{eqnarray}
\text{Im}(E)=\sum_{n}|t_n|(|\beta_M|^n-|\beta_M|^{-n})\sin(n\theta+\phi_n)=\frac{2a_1}{L}\sum_{n}n|t_n|\sin(n\theta+\phi_n)+O(\frac{1}{L^2})=0,
\end{eqnarray}
which has to be satisfied for each order of $1/L$ when $L\gg 1$. Even in the leading order, 
only limited number of $\theta$ could meet this condition, thus the  continuous spectrum only contains few isolated small spots (due to limited number of $\theta$), which apparently can not produce a meaningful  continuous spectrum. Therefore, real-valued  continuous spectrum requires unit modulus for $\beta_M$ ($a_n=0$), i.e., the eigenstate is extended.
\\

\textbf{PBC case}. The PBC case follows the same logic but is a little bit more complicated. A general Hamiltonian in a closed chain is given by $H^{(p)}=\sum_{n=1}^M h^{(p)}_n+V$ where $V$ is the same as the OBC case, $h^{(p)}_n$ represents the hopping terms, containing some terms bridging the two ends of the chain 
\begin{eqnarray}
h^{(p)}_n=\sum_{i=1}^{L-n}t_n|i\rangle\langle i+n|+ t_n^\ast |i+n\rangle\langle i|+\sum_{i=L-n+1}^L t_n|i\rangle\langle i+n-L|+t^\ast_n|i+n-L\rangle\langle i|.
\end{eqnarray}
Similar as the OBC case, the eigenvalues and eigenstates can be solved by taking the same ansatz $|\beta\rangle$ such that
\begin{eqnarray}
h^{(p)}_n|\beta\rangle=(t_n\beta^n+t_n^\ast \beta^{-n})|\beta\rangle+t_n(1-\beta^{L-n})\sum_{l=1}^n \beta^l|L-(n-l)\rangle+t_n^\ast \beta^{-n}(\beta^{L}-1)\sum_{l=1}^n\beta^l|l\rangle,
\end{eqnarray}
and the operation produced by non-Hermitian boundary stay the same as Eq.~\eqref{eq:findeigenstate}.
By linearly constructing the same wavefunciton in Eq.~\eqref{eq:wavefunction} and acting the Hamiltonian on it, we obtain
\begin{eqnarray}
H^{(p)}|\psi\rangle=E|\psi\rangle+B^{(p)}
\end{eqnarray}
where
\begin{eqnarray}
B^{(p)}=\sum_{i=1}^M \sum_{n=i}^M\sum_{l=1}^{2M}[ c_lt_n^\ast(\beta_l^{L-n+i}-\beta_l^{i-n})|i\rangle+c_lt_n(\beta_l^{i}-\beta_l^{L+i})|L-n+i\rangle]+\sum_{l=1}^{2M}\sum_{i\in\partial\Omega}c_lv_i\beta_l^i|i\rangle+\sum_{l=1}^{2M}\sum_{i,j\in\partial\Omega}c_lv_{ij}\beta_l^j|i\rangle.\nonumber\\
\end{eqnarray}
Different from OBC case, the extra term $B^{(p)}$ contains a term $\sim \beta_l^L$ on both boundary. As before, eliminating $B^{(p)}$ gives the boundary equation $F^{(p)}(\{\beta_i\})\cdot(c_1,c_2,\cdots,c_{2M})^T=0$ with
\begin{eqnarray}
F^{(p)}(\{\beta_i\})=\left(
\begin{array}{cccccc}
    \beta_1^Lf^{(a)}_1(\beta_1)+f^{(b)}_1(\beta_1)&\cdots &\beta_{2M}^Lf^{(a)}_1(\beta_{2M})+f^{(b)}_1(\beta_{2M})\\ \vdots&\vdots&\vdots\\
    \beta_1^Lf^{(a)}_M(\beta_1)+f^{(b)}_M(\beta_1)&\cdots &\beta_{2M}^Lf^{(a)}_M(\beta_{2M})+f^{(b)}_M(\beta_{2M})\\
    \beta_1^Lg^{(a)}_1(\beta_1)+g^{(b)}_1(\beta_1)&\cdots &\beta_{2M}^Lg^{(a)}_1(\beta_{2M})+g^{(b)}_1(\beta_{2M})\\ \vdots&\vdots&\vdots\\
    \beta_1^Lg^{(a)}_M(\beta_1)+g^{(b)}_M(\beta_1)&\cdots &\beta_{2M}^Lg^{(a)}_M(\beta_{2M})+g^{(b)}_M(\beta_{2M})
\end{array}
\right),
\end{eqnarray}
where $f^{(a,b)}_i(\beta_j)$, $g^{(a,b)}_i(\beta_j)$ are polynomials of $\beta_j$ with highest order less than the hopping range $M$. Since real eigenvalues enforce $\beta_j=1/\beta_{2M-j+1}^\ast$, it is obvious that $|\beta_{i\geq M+1}|\geq 1$ and $|\beta_{i\leq M}|\leq 1$. Therefore, the leading and sub-leading term in the boundary condition are $(\beta_{2M}\cdots \beta_{M+2}\beta_{M+1})^L$ and $(\beta_{2M}\cdots \beta_{M+2})^L$. For a large system with $L\gg 1$, it requires that these two term could compete with each other to possibly satisfy the boundary equation. Hence, $|\beta_{M+1}|^L\sim 1$ is required, which is also equivalent to $|\beta_M|^L=1/|\beta_{M+1}^\ast|^L\sim 1$. Now, the situation resembles the OBC case, and by the same logic we can conclude that the eigenstates always distribute uniformly as long as the eigenvalue is real.
\\

\textbf{For complex eigenvalues}. On the contrary, if the spectrum takes complex value,  $|\beta|\neq 1$ is required for the complex eigenvalues because $\text{Im}[E(\beta)]_{|\beta|=1}=0$. This indicates that the decaying eigenstate (scale-free localization) is necessary for a system with complex eigenvalues and this can be regarded an accompanying effect of PT breaking induced by a boundary non-Hermitian term in a Hermitian system.

\subsection{Eigenstates for PT-broken phase}

Now we know that the eigenstates in the PT-broken phase cannot be extended because complex eigenvalues requires $|\beta|\neq 1$. However, it is still necessary to confirm the appearance of scale-free localization in the PT-broken phase. In the discussion above, a key step for justification is to show $|\beta_M|^L\sim 1$, which heavily relies on the fact that the solution for $\beta$ appears in pairs, i.e., $(\beta,1/\beta^\ast)$, especially for the discussion of OBC case. 
Here, we emphasize that if the system respect PT symmetry, then its eigenvalues exist in pairs $(E,E^\ast)$. According to form of eigenvalues in Eq.~\eqref{eq:generalenergy}, it is clear that the solution for $\beta$ also form pairs $(\beta,1/\beta^\ast)$ for a PT symmetric system. Therefore, all the deduction for the emergence of scale-free localization works perfectly to PT symmetric phase and the complex eigenvalues in the PT-broken phase is compatible with it.


\begin{figure}
\begin{tabular}{cc}
\includegraphics[width=0.8\linewidth]{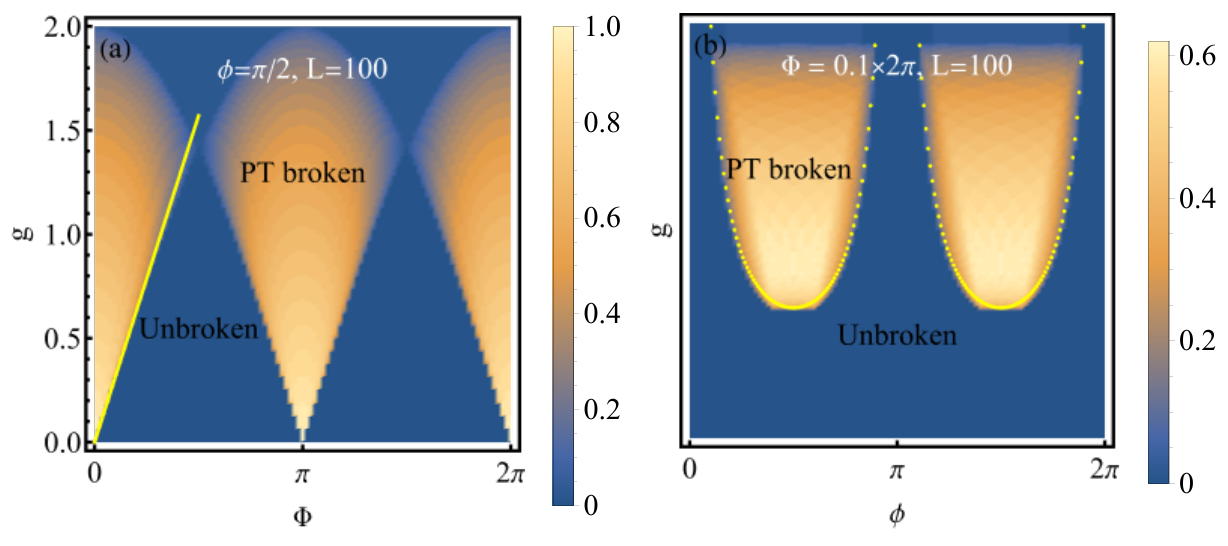}\\
\includegraphics[width=0.8\linewidth]{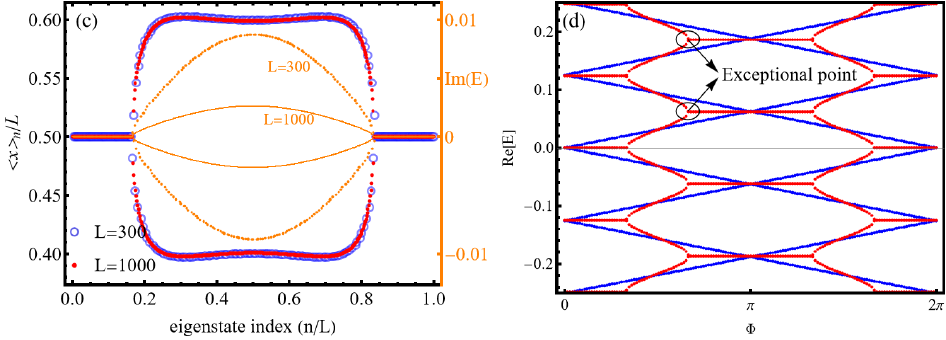}
\end{tabular}
 \caption{(a),(b) Phase diagram for a periodic chain under a local non-Hermitian perturbation. The color map shows the proportion of complex eigenvalues $P_{\text{com}}$, which serves as an indicator of PT symmetry breaking. The horizontal axis is for $\Phi$ in (a), and $\phi$ in (b). (c) The mean position $\langle x\rangle_n$ (left ticks, black) and imaginary part of eigenvalues (right ticks, orange), where $g=1,\phi=\pi/2$ are used, and eigenstate index is arranged in the ascending order of eigenvalue real part. (d) The real part of spectrum as a function of total flux $\Phi$. The parameters values are:  blue, $g=0$; red, $g=1$, $\phi=\pi/2$. The red lines contain exceptional points (EPs) induced by non-Hermitian perturbation. }\label{fig:phasePBC}
\end{figure}

\section{PT breaking induced by local non-Hermitian impurity under PBC}

In this section, we analyze the PT breaking in a PBC model by exploring its eigenvalues and build up a phenomenological frame to understand the behavior intuitively. We consider the following model $H=H_p+V$ with
\begin{eqnarray}\label{eq:simplechain}
H_p&&=\sum_{j=1}^{L-1} (te^{i\theta}|i\rangle\langle i+1|+te^{-i\theta} |i+1\rangle\langle i|)+t e^{-i\theta} |1\rangle \langle L|+te^{i\theta}|L\rangle\langle 1|,\nonumber\\
V&&=ge^{i\phi}|1\rangle\langle 1|+ge^{-i\phi}|L\rangle\langle L|.
\end{eqnarray}
Here, $H_p$ describes a periodic chain enclosing a magnetic flux $\Phi=L\theta$, and $V$ is generally a non-Hermitian potential (except for $\phi=0,\pi$).
For convenience, we perform similarity transformation $U=\sum_{j=1}^L e^{i\theta j}|j\rangle \langle j|$ to convert the Hamiltonian to $\Tilde{H}=\Tilde{H}_p+\Tilde{V}$ with 
\begin{eqnarray}
\Tilde{H}_p&&=U^\dagger H_p U= \sum_{j=1}^{L-1} (t|i\rangle\langle i+1|+t |i+1\rangle\langle i|)+t e^{-i\theta L} |1\rangle \langle L|+te^{i\theta L}|L\rangle\langle 1|,\nonumber\\
\Tilde{V}&&=V.
\end{eqnarray}
Here, $\Tilde{H}$ shares the same eigenvalues with $H$ and the corresponding eigenstates are connected by $U$ through $|\psi\rangle =U|\Tilde{\psi}\rangle$. After the transformation, $\Tilde{H}_p$ is a periodic function of $\theta$, i.e., $\Tilde{H}_p(\theta+2\pi n/L)=\Tilde{H}_p(\theta)$ with $n\in\mathbbm{Z}$. The eigenvalue of $\Tilde{H}_p$ is
\begin{eqnarray}\label{eq:PBCbands}
E_n=2\cos(\frac{2\pi n}{L}+\theta),\qquad 
n=1,2,\cdots, L.
\end{eqnarray}
For $\theta= m\pi/L$, the eigenvalues are doubly degenerate as $\cos(2\pi n/L+m\pi/L)=\cos[2\pi(L-n-m)/L+m\pi/L]$ $(m,n\in \mathbbm{Z})$ where the argument of cosine is defined for modulation with respect to $2\pi$. From the phase diagram in the main text, the PT breaking threshold vanishes at these values.

The PT breaking can be quantified by the proportion of complex eigenvalues $P_{\text{com}}=n_{\text{com}}/L$ with $n_{\text{com}}$ being the number of complex eigenvalues. As shown in  Fig.~\eqref{fig:phasePBC} (a), (b), phase diagrams are given with respect to different parameters.
Fig.~\ref{fig:phasePBC} (c) shows that the complex eigenvalues exhibit scale-free localization (characterized by $\langle x\rangle_n$), confirming our predictions.
The real part of spectrum in Fig.~\ref{fig:phasePBC} (d) shows that the PT breaking takes place by deforming nearest real energy levels into complex through an exceptional point (EP), at which both eigenvalues and eigenstates coalesce. This stimulates an intuitive understanding by projecting the Hamiltonian into two-level subspaces. In the following, we firstly solve the boundary condition to show that the emergence of scale-free localization is inevitable in the PT-broken phase, which is consistent with the prediction in the main text, then we create a two-level subspace effective theory to gain an intuitive understanding about the PT breaking.

\subsection{Solving boundary condition}

\begin{figure} \includegraphics[width=0.8\linewidth]{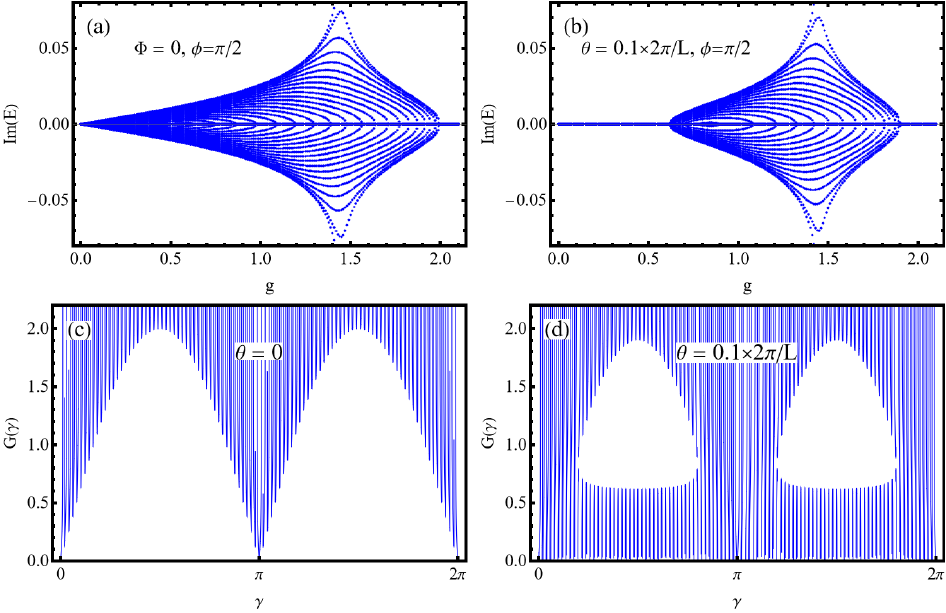}
 \caption{PT breaking for examples with $\phi=\pi/2$. (a,b) give imaginary energy as a function of non-Hermitian parameter $g$. (c,d) plot the function $G(\gamma)$ defined in Eq.~\eqref{eq:unitsolution}, corresponding the two cases in (a,b), respectively.}\label{fig:simplemodel}
\end{figure}

Like before, the characteristic equation of $\Tilde{H}$ is given by $t(\beta+\beta^{-1})=E$ and the eigenstate is $|\Tilde{\psi}(\beta)\rangle=c_1 |\beta\rangle+c_2 |\beta^{-1}\rangle$, such that
\begin{eqnarray}
\Tilde{H}|\Tilde{\psi}(\beta)\rangle=E|\Tilde{\psi}(\beta)\rangle&&+\big(c_1[t(e^{-i\theta L}\beta^L-1)+ge^{i\phi} \beta]+c_2 [t(e^{-i\theta L}\beta^{-L}-1)+g e^{i\phi} \beta^{-1}]\big)|1\rangle\nonumber\\
&&+\big( c_1[t(e^{i\theta L}\beta-\beta^{L+1})+g e^{-i\phi}\beta^L]+c_2[t(e^{i\theta L}\beta^{-1}-\beta^{-(L+1)})+g e^{-i\phi}\beta^{-L}]
\big)|L\rangle.
\end{eqnarray}
The vanishing of the tail in the r.h.s. of the equation above requires $\det (F)=0$ with
\begin{eqnarray}
F=\left(\begin{array}{cc}
 t(e^{-i\theta L}\beta^L-1)+ge^{i\phi} \beta&t(e^{-i\theta L}\beta^{-L}-1)+g e^{i\phi} \beta^{-1} \\
  t(e^{i\theta L}\beta-\beta^{L+1})+g e^{-i\phi}\beta^L& t(e^{i\theta L}\beta^{-1}-\beta^{-(L+1)})+g e^{-i\phi}\beta^{-L} 
\end{array}\right),
\end{eqnarray}
which is the boundary condition,  determining the solutions for $\beta$. As analyzed before, the emergence of imaginary solution (PT breaking) is equivalent to existing solution $|\beta|\neq 1$. In other words, unitary solutions fails in the PT-broken phase, so one can plug in $\beta=e^{i\gamma}$ ($\gamma\in [0,2\pi]$) and detect the parameter region where the ansatz can not give meaningful results.  By substituting the solution ansatz $\beta=e^{i\gamma}$ into the boundary condition leads to 
\begin{eqnarray}
(\frac{g}{t})^2\sin[\gamma(L-1)]-2\frac{g}{t}\cos\phi\sin(\gamma L)+2[\cos(\gamma L)-\cos(\theta L)]\sin\gamma=0,
\end{eqnarray}
which can be used to solve $\theta$ for a given $g$,  or express $g/t$ inversely in terms of $\theta$, i.e.,
\begin{eqnarray}\label{eq:unitsolution}
\frac{g}{t}=G(
\gamma)=\frac{\cos\phi\sin(\gamma L)\pm\sqrt{\cos^2\phi\sin^2(\gamma L)-2\sin[\gamma(L-1)][\cos(\gamma L)-\cos(\theta L)]}}{\sin[\gamma(L-1)]}.
\end{eqnarray}
As shown in Fig.~\ref{fig:simplemodel} (c,d), in the PT-broken phase, there is certain region for $\gamma$ where $G(\gamma)$ does not always give real value, i.e., the unitary ansatz can not give all solutions in certain range of $G(\gamma)$, equivalent to say $g$, implying PT symmetry is broken in this regime. Especially, compared with Fig.~\ref{fig:simplemodel} (a,b), this range of $G(\gamma)$ perfectly matches the range for $g/t$ in which PT symmetry is broken.
\\

In the PT-broken phase, the solutions are given by the ansatz $|\beta|=e^{\delta/L}$ which is generally correct up to the order $1/L$. By substituting $\beta=e^{\delta/L+i\gamma}$ into $\det F$, we obtain $\det F=f_R+i f_I$ with
\begin{eqnarray}
f_R&&=2\sinh{\delta}[2t^2\sin(\gamma L)\sin\gamma-g^2\cos(\gamma(L-1))+2gt\cos\phi\cos(\gamma L)]+O(\frac{1}{L}),\nonumber\\
f_I&&=2\cosh{\delta}[-g^2\sin(\gamma(L-1))+2gt\cos\phi\sin(\gamma L)-2t^2\cos(\gamma L)\sin\gamma]+4t^2\cos(\theta L)\sin\gamma+O(\frac{1}{L}).
\end{eqnarray}
At the leading order, the boundary condition can give satisfactory value for $\gamma$ in the region where the unitary solution fails from $f_R=0$; the value of $\delta$ can be subsequently solved by substituting the resultant value of $\gamma$ into $f_I=0$.
\\

\textbf{Robust  continuous spectrum under OBC}. As a comparison, we also briefly analyze the open boundary case, $H=H_o+V$
\begin{eqnarray}
H_o&&=\sum_{j=1}^{L-1} (te^{i\theta}|i\rangle\langle i+1|+te^{-i\theta} |i+1\rangle\langle i|)
\end{eqnarray}
where the phase factor $\theta$ can be always removed by the similarity transformation $U$ introduced before. By following the same procedure, the boundary condition here is reduced to 
\begin{eqnarray}\label{eq:nearestcase}
\det\left(\begin{array}{cc}
    g e^{i\phi}\beta-t & g e^{i\phi}\beta^{-1}-t \\
    g e^{-i\phi}\beta^L-t\beta^{L+1}&  g e^{-i\phi}\beta^{-L}-t\beta^{-(L+1)}
\end{array}\right)=0.
\end{eqnarray}
As we discussed before, the existence of complex eigenvalues is equivalent to existing solution $\beta$ with non-unitary module. If we assume a solution $|\beta|>1$ exist and $|\beta|^L\gg |\beta|^{-L}$, then the coefficient of leading order $\beta^L$ on the l.h.s. of equation above has to vanish, i.e., $(g e^{-i\phi}-t\beta)(g e^{i\phi}-t\beta)=0$, which gives
\begin{eqnarray}
\beta=\frac{g}{t}e^{\pm i\phi}.
\end{eqnarray}
This suggests that only when $|g|>t_1$, there could exist two modes with $|\beta|>1$. These modes are not  continuous-spectrum PT-broken modes because of its size-independent module and limited number. Another possibility is that $\beta^L\sim \beta^{-L}$, by which we can assume $\beta=e^{i\gamma+\delta/L}$ [correct to the order O$(1/L)$]. Plugging this expression into Eq.~\eqref{eq:nearestcase} and keeping leading order results in
\begin{eqnarray}
e^{2\delta}=\frac{g^2-2gt\cos\phi e^{-i\gamma}+ t^2e^{-i2\gamma}}{g^2-2gt\cos\phi e^{i\gamma}+ t^2e^{i2\gamma}} e^{-i\gamma 2(L-1)}.
\end{eqnarray}
Note that $|\text{r.h.s.}|=1$ and  $\delta\in \mathbbm{ R}$, so the only possible solution is $\delta=0$, i.e., the  continuous-spectrum PT breaking can not happen. Since there are only two equal energy points for any energy in the band structure of the model with only nearest hopping, this result is indeed consistent with our statement on  continuous-spectrum PT breaking condition. From the simple analysis, we conclude that a system can response much differently to a local non-Hermitian perturbation under PBC and OBC.

\subsection{Effective theory}

In this part, we construct an effective theory from phenomenological aspect to gain more intuitive understanding about the PT transition behavior. As shown in the right column of Fig.~\ref{fig:phasePBC} (d) in main text, the local non-Hermitian impurity can couple two nearest energy levels to form  2$nd$ order EP and break the PT symmetry. So we attempt to build an effective theory to describe each two-level subsystem by using projecting method. For two nearest levels (even degenerate levels) with energy $E_{n_1}, E_{n_2}$, assume $|E_{n_1}-E_{n_2}|\ll\Gamma=\min_{m\neq n_{1/2}}\{|E_{n_{1/2}}-E_{m}|\}$, one can project the eigen-equation into the subspace spanned by $n_1$, $n_2$ modes via projector $P_{n_1n_2}=|\psi_{n_1}\rangle\langle \psi_{n_1}|+|\psi_{n_2}\rangle\langle\psi_{n_2}|$, namely
\begin{eqnarray}
\big[P_{n_1n_2}HP_{n_1n_2}-E-P_{n_1n_2}HQ(QHQ-E)^{-1}QHP_{n_1n_2}\big]\psi_P=0
\end{eqnarray}
where $Q=\mathbbm{1}-P_{n_1n_2}$ and $\psi_P$ stands for the wavefunction in the subspace. Here, $H=H_0+V$ with $H_0|\psi_n\rangle=E_n|\psi_n\rangle$ and $V$ only containing nonzero components near the boundary. Since the purpose of the projection is to explore the bands deformed from $E_{n_{1/2}}$, we focus on the energy scale $E_{n_2}\leq E\leq E_{n_1}$ and $||V||\simeq |E_{n_1}-E_{n_2}|$, and consider that $P_{n_1n_2}HQ=P_{n_1n_2}VQ$, $QHP_{n_1n_2}=QVP_{n_1n_2}$, then
\begin{eqnarray}
(E-QHQ)^{-1}=[G_Q^{-1}(1-G_Q V)]^{-1}=\sum_{n=0}^\infty (G_QV)^n G_Q
\end{eqnarray}
where $G_Q=(E-QH_0Q)^{-1}=\sum_{m\neq n_1,n_2}|\psi_m\rangle\langle\psi_m|/(E-E_m)$. Therefore,
\begin{eqnarray}
P_{n_1n_2}HQ(QHQ-E)^{-1}QHP_{n_1n_2}=P_{n_1n_2}VQ\sum_{n=0}^\infty (G_QV)^n G_QQVP_{n_1n_2}\sim \mathcal O(\frac{||V||}{\Gamma}),
\end{eqnarray}
which can be approximately dropped out under the assumption $||V||\ll \Gamma$. This approximation should works extremely well near the gapless point. The effective Hamiltonian is approximately given by
\begin{eqnarray}\label{eq:effectiveHamiltonian}
h_{eff}\approx P_{n_1n_2}HP_{n_1n_2}=\bar{\epsilon}_{12}+\Delta_{12}\sigma_z+P_{n_1n_2}VP_{n_1n_2},
\end{eqnarray}
where $\bar{\epsilon}_{12}=\big(E_{n_1}+E_{n_2}\big)/2$ and  $\Delta_{12}=\big(E_{n_1}-E_{n_2}\big)/2$, and the effective Hamiltonian is valid for energy scale $E_{n_2}\leq E\leq E_{n_1}$.
\\

In the following, we will try to utilize the effective theory to study the condition for PT breaking happened in the PBC model above. For convenience, we assume that the origin of the coordinates locate at the middle of the chain such that $x_L=-x_1=(L-1)/2$. The unperturbed eigenstate is plane wave $|\psi_m\rangle=\frac{1}{\sqrt{L}}\sum_{j}e^{ik_mx_j}|j\rangle$, which gives 
\begin{eqnarray}
\langle\psi_m|V|\psi_n\rangle=\frac{2g}{L}\cos[(k_m-k_n)\frac{L-1}{2}+\phi].
\end{eqnarray}
Here, for simplicity, we focus on $0\leq \theta< 0.25(2\pi)/L$ where the spectrum has degeneracy at $\theta=0$, and the PT breaking could possibly take place between $k_n$ and $k_{L-n}$, so the involved effective Hamiltonian is given by
\begin{eqnarray}
h_{eff}=\bar{\epsilon}_{12}+\Delta_{12}\sigma_z+\frac{2g}{L}[\cos\phi\sigma_0-(-1)^L\cos(k_n)\cos\phi\sigma_x-i(-1)^L\sin(k_n)\sin\phi\sigma_y]
\end{eqnarray}
where $\sigma_i$ ($i=x,y,z$) stands for the Pauli matrices, $\Delta_{12}=-2\sin k_n\sin\theta$, and we used $(k_n-k_{L-n})\frac{L-1}{2}=\pi(2n-L)+\pi-k_n$.
Solving the eigenvalues, one sees that the PT symmetry is broken when $(2\sin k_n\sin\theta)^2+(\frac{2g}{L})^2(\cos^2 k_n\cos^2\phi-\sin^2 k_n\sin^2\phi)<0$,
which gives the threshold $g_c$
\begin{eqnarray}\label{eq:generalthreshold}
\Big(\frac{g_c/t}{L}\Big)^2=\sin\theta\min_{k_n,\text{positive}}\Big(\frac{-2\sin^2 k_n}{\cos2k_n+\cos2\phi}\Big).
\end{eqnarray}
Here, $\min_{k_n,\text{positive}}(\cdots)$ means minimizing the function in the brackets for all possible $k_n$. In Fig.~\ref{fig:phasePBC}, the yellow line in (a) and the dots in (b) are obtained from this condition, which agrees well with the numerical results.  Specifically, in (a) with $\phi=\pi/2$, the PT transition threshold is
\begin{eqnarray}\label{eq:specialthreshold}
g_c/t=L\sin\theta\approx\Phi,
\end{eqnarray}
corresponding to the yellow line. 
\\

\textbf{Emergent (approximated) chiral symmetry and stability of real spectrum}---From the phase diagram Fig.~\ref{fig:phasePBC}(a) in the main text, we find when $\theta\simeq 0.25\times 2\pi/L$ and $\phi=\pi/2$, the PT-broken phase is absent. This is because the gap between neighboring energy levels are almost equal, which leads to the break down of the two-level effective Hamiltonian approach discussed above. To understand the stability of PT symmetric phase in this situation, one needs to construct an effective  theory including more relevant energy levels. The evenly spaced gaps around $\theta\simeq 0.25\times 2\pi/L$ can be checked as the below:  for $\theta\leq 0.25\times 2\pi/L$, the eigenvalues for three nearest eigenvalues (labeled by $-1,0,+1$) read $\epsilon_{-1}=\cos[\frac{2\pi (n-1)}{L}+\eta\frac{2\pi}{L}]$, $\epsilon_0=\cos[\frac{2\pi (L-n)}{L}+\eta\frac{2\pi}{L}]$ and $\epsilon_{+1}=\cos[\frac{2\pi n}{L}+\eta\frac{2\pi}{L}]$ where $n\leq L/2$ ($n\in\mathbbm{Z}_+$) and $\eta\leq 0.25$ (with $\theta=\eta 2\pi/L$).  It is easy to check up to the order $1/L$
\begin{eqnarray}
\epsilon_{-1}-\epsilon_0=\sin(\frac{2\pi n}{L})(1-2\eta)\frac{2\pi}{L}+O(\frac{1}{L^2}),\qquad
\epsilon_{0}-\epsilon_{+1}=\sin(\frac{2\pi n}{L})(2\eta)\frac{2\pi}{L}+O(\frac{1}{L^2}),
\end{eqnarray}
and $\epsilon_{-1}-\epsilon_0\simeq\epsilon_0-\epsilon_{+1}$ when $\eta=0.25$. 
\\

Here, we put the analysis of spectrum stability to a more general ground. Let us consider a non-degenerate spectrum, in which a given eigenvalue $E_n$ is symmetrically surrounded by several approximately evenly spaced neighboring eigenvalues. With relabeling $E_0=E_n$, the subset of eigenvalues near $E_0$ is given by $(E_{-w},E_{-(w-1)},\cdots,E_0,\cdots E_{w-1}, E_{w})$ where $|E_{l+1}-E_{l}|\simeq \delta E$ (where $-w\leq l< w$). Now, when we concentrate on a energy scale $E\simeq E_0$ and a moderate perturbation  $||V||\simeq\delta E$, an effective Hamiltonian can be constructed by performing projection into the space spanned by the subset.  The projection scheme for the two-level case still works by neglecting a correction of order $O(||V||/\Gamma)\simeq O(1/w)\ll 1$, where $\Gamma=w\delta E$. Up to a constant $E_0$, the effective Hamiltonian is 
\begin{eqnarray}\label{eq:multibandeffective}
h_{eff}=P_w HP_w=\text{diag}\{-w,-(w-1),\cdots, 0, \cdots, (w-1),w\}\delta E+\mathcal V
\end{eqnarray}
with $P_w=\sum_{l=-w}^w|\psi_{l}\rangle\langle \psi_l|$ and $\mathcal V=P_w VP_w$. Note this effective Hamiltonian only works well for $-\delta E\leq E\leq\delta E$. Here, if the boundary term induced coupling satisfy inversion asymmetry 
\begin{eqnarray}
	\mathcal V_{i,j}=-\mathcal V_{-i,-j},\qquad 
	i,j\in (-w,\cdots,0,\cdots,w),
\end{eqnarray}
the full effective Hamiltonian will be inversion asymmetric 
\begin{eqnarray}\label{eq:chiral}
	\Tilde{\mathcal P} h_{eff}\Tilde{\mathcal P}=-h_{eff},
\end{eqnarray}
where $\Tilde{\mathcal P}$ is inversion operation of eigenvalue index with respect to the middle level, $\Tilde{\mathcal P}: j\rightarrow -j$. This actually is a chiral symmetry and it enforces the eigenvalues to appear in pairs with opposite sign, so that the middle level is zero, i.e., $E=0$. This demonstrates that the boundary term can not deform the interested level effectively, so it can not merges into a EP with other levels and break the PT symmetry.

In the model studied before, the modes $k_n$ and $k_{L-n}$ are degenerate in the absence of phase of hopping parameter, i.e., $\theta=0$. Adding the flux lifts the degeneracy by shifting the effective momentum by $\theta$, i.e., $k_n\rightarrow k_n+\theta$. For $0< \theta<\pi/L$, the bands in Eq.~\eqref{eq:PBCbands} is ordered by 
\begin{eqnarray}
\cdots>E(k_{n-1}+\theta)>E(k_{L-n}+\theta)>E(k_n+\theta)>E(k_{L-(n+1)}+\theta)>\cdots.
\end{eqnarray}
At $\theta=0.25\times \pi/L$, these levels are approximately spaced evenly if only a few  eigenvalues near $E(k_n+\theta)$ is counted, i.e., $w\ll L$. The corresponding eigenstates are irrelevant with $\theta$ due to translational symmetry.
We regard $E(k_n+\theta)$ as the reference band and index the subspace by a ascending order of eigenvalues, then the eigenstate with new index can be attached with a consistent label
\begin{eqnarray}
|\psi_l\rangle=\frac{1}{\sqrt{L}}\sum_{j=1}^L e^{iK_l j}|j\rangle, \qquad (l=-w,\cdots w)
\end{eqnarray}
where
\begin{eqnarray}
&&K_0=k_n, \nonumber\\
&&K_{-1}=k_{L-{n+1}},\qquad
K_{-2}=k_{n+1}, \cdots\nonumber\\
&&K_1=k_{L-n},\qquad
K_{2}=k_{n-1},\cdots.
\end{eqnarray}
With these labeling,
the potential projection at $\phi=\pi/2$ reads
\begin{eqnarray}
	\mathcal V_{i,j}=-\frac{2g}{L}=\sin[(K_i-K_j)\frac{L-1}{2}]
\end{eqnarray}
where we used $x_1=x_L=-\frac{L-1}{2}$.
This projected perturbation is asymmetric upon switching the index $i,j$. Combing with the asymmetric diagonal terms (upon switching index) in Eq.~\eqref{eq:multibandeffective}, the effective Hamiltonian respects a chiral symmetry Eq.~\eqref{eq:chiral}. Therefore, the resultant reference band is untouched, namely the PT symmetry for eigenvalue of our concern ($E_n$)  is not broken.

\section{Boundary perturbation induced PT breaking under OBC}

In this section, we give more details about the parity analysis of the OBC model to lay down the foundation of effective sub-band theory presented in the main text. The model is recapped here 
\begin{eqnarray}\label{eq:OBCmodel}
	H_0=\sum_{j=1}^{L-1}t_1|j\rangle\langle j+1| + \sum_{j=1}^{L-2}t_2|j\rangle\langle j+2| +H.c.
\end{eqnarray}
where $t_1$, $t_2$ are real hopping strength.
And we consider two kinds of non-Hermitian boundary potential
\begin{eqnarray}\label{eq:OBCperturbation}
	V_1&&=i g ( |1\rangle\langle 1|- |L\rangle\langle L|),\nonumber\\
	V_2&&= g(|1\rangle\langle 2|+ |L\rangle\langle L-1|),
\end{eqnarray}
where $V_1$ is discussed in main text, $V_2$ will be considered later as a comparison with $V_1$.

\begin{figure}
\begin{tabular}{cc}
      \includegraphics[width=0.8\linewidth]{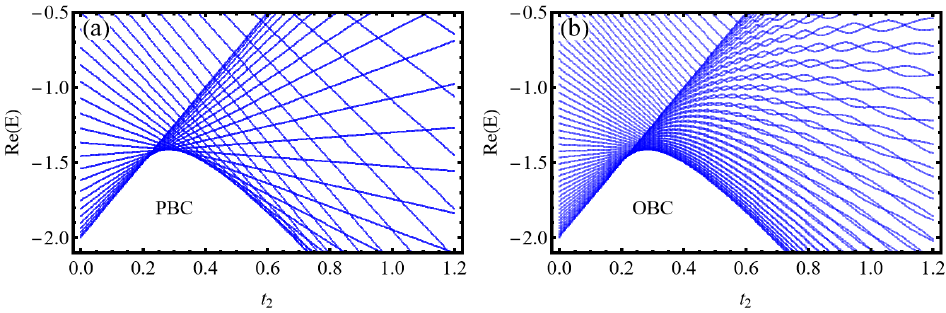}\\
       \includegraphics[width=0.8\linewidth]{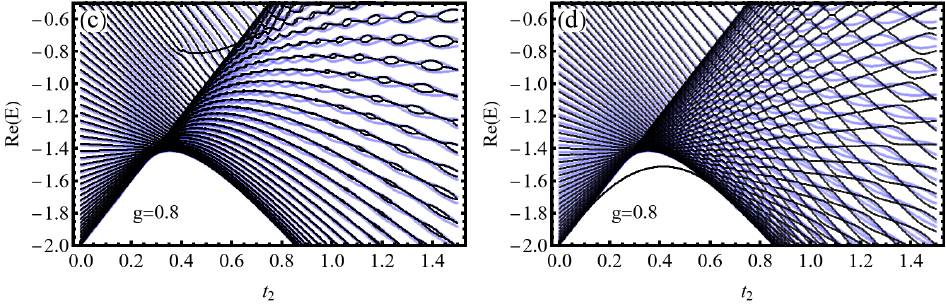}
\end{tabular}
 \caption{Spectrum for models described by Eq.~\eqref{eq:OBCmodel} and~\eqref{eq:OBCperturbation}. (a,b): the PBC and OBC spectrum without non-Hermitian boundary terms. (c,d): spectrum (black) with boundary $V_1$  and $V_2$ overlaying on top of the OBC spectrum (blue). Here, the lines deviate from typical shape of  continuous spectrum reflect the bound states exponentially localized near the perturbation. All plots are made with $t_1=1$ and $L=100$.}\label{fig:staticreal}
\end{figure}

The real part of spectrum is depicted in Fig.~\ref{fig:staticreal}, where the spectrum of $H_0$ with PBC and OBC is given in (a) and (b), respectively, and the spectrum of model with perturbation $V_1$, $V_2$ are individually given in (c), (d) by overlaying them on top of OBC spectrum of $H_0$. By comparing (a),(b), some degeneracies in the PBC are lifted under OBC while others are preserved. This is because $H_0$ has an inversion symmetry, i.e., its eigenstates fall into even or odd parity sector. Under PBC, the eigensates of opposite parity are fully degenerate. Suppose the $L=2M+1$ (it is similar for even number of sites), the eigenvalue and eigenstate are  respectively given by $E(k)=2t_1\cos(k)+2t_2\cos(2k)$ and  $\frac{1}{\sqrt{L}}e^{i kx}$ with $k=\frac{2\pi n}{L},$ and $n=0,\pm 1,\cdots,\pm M$. It is obvious that $\frac{1}{\sqrt{L}}e^{\pm ikx}$ (for $k\neq 0$) are degenerate so that they can be recombined to form eigenstate with opposite parity
\begin{eqnarray}
\psi^{(even)}_q(x)=\sqrt{\frac{2}{L}}\cos (qx),\qquad  
\psi^{(odd)}_q(x)=\sqrt{\frac{2}{L}}\sin (qx),
\end{eqnarray}
and the domain of $q$ is shrank by half, $q=\frac{2\pi m}{L}$ and $m=0,1,2,\cdots, M$, where $k$ is replaced by $q$ to distinguish their domains. Here, we assume $x=-M,-(M-1),\cdots,-1, 0,1,\cdots, M-1,M$ to show the parity of eigenbasis in the most apparent way. Besides the parity degeneracy, after $t_2$ reaching certain value (when $t_2>t_1/4$, four equal energy points become possible), there are many accidental degeneracies between different $q$ modes in the PBC spectrum, which are fragile and a gap can be easily open if there is no other symmetry protection.

\begin{figure}
     \includegraphics[width=1\linewidth]{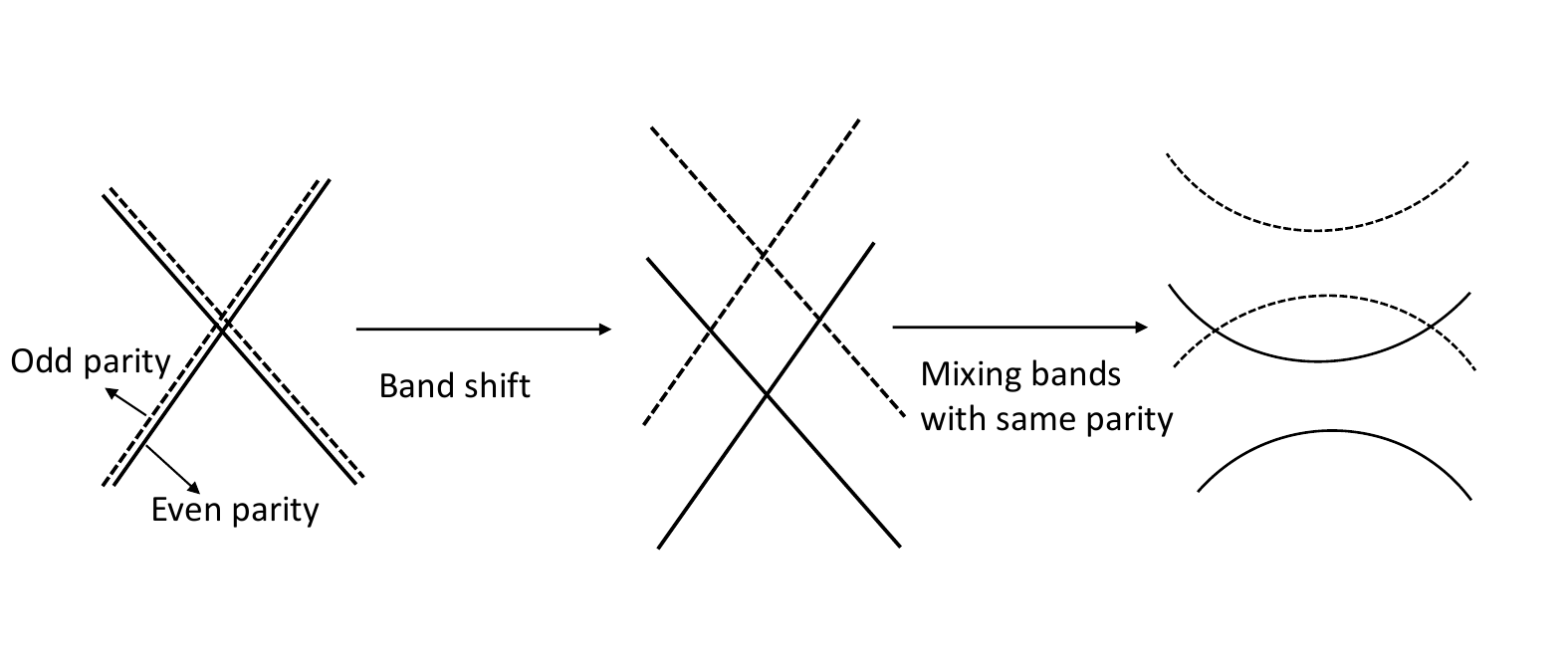} 
\caption{Schematic plot for gap opening and band shift between different bands with opposite parity.  }\label{fig:schematic}
\end{figure}

The OBC spectrum can be obtained by adding  boundary terms to PBC Hamiltonian, i.e., $H_{\text{OBC}}=H_{\text{PBC}}+V_{\text{OBC}}$ with 
\begin{eqnarray}
V_{\text{OBC}}=-t_1\big(|1\rangle
\langle L|+|L\rangle\langle 1|\big)-t_2\big(|1\rangle
\langle L-1|+|L-1\rangle\langle 1|+|2\rangle
\langle L|+|L\rangle\langle 2|\big).
\end{eqnarray}
 It is evident that $V_{\text{OBC}}$ preserves the inversion symmetry so it cannot mix any two PBC eigenstates of opposite parity.  The accidental degeneracy between different $q$ modes in PBC spectrum is actually 4-fold degenerate because each $q$ mode is 2-fold degenerate with respect to parity. \textit{$V_{\text{OBC}}$ can only open a gap between eigenstates of same parity, the gapless points between opposite parity can be only shifted, as shown schematically in Fig.~\ref{fig:schematic}}. As a result, the full degeneracy between opposite parity (belonging to the same $q$ mode) is lifted due to the band shift induced by $V_{\text{OBC}}$, which explains the absence of eigenvalue degeneracy under OBC in the regime $t_2<t_1/4$; while the gapless point between modes of opposite parity (belonging to different $q$ modes) is preserved.  So in the Fig.~\ref{fig:staticreal} (b),  the braid shape bands with gapless points belong to opposite parity. This parity categorization of energy levels is crucial for the effective theory discussed in the main text for perturbation $V_1$, and will be utilized as well in analyzing the effect of $V_2$ in the following.

\begin{figure}
\begin{tabular}{cc}
\includegraphics[width=0.8\linewidth]{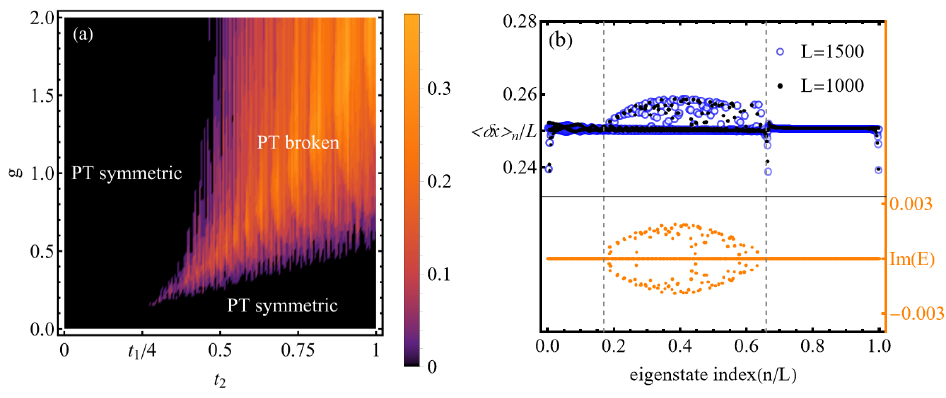}\\
\includegraphics[width=0.75\linewidth]{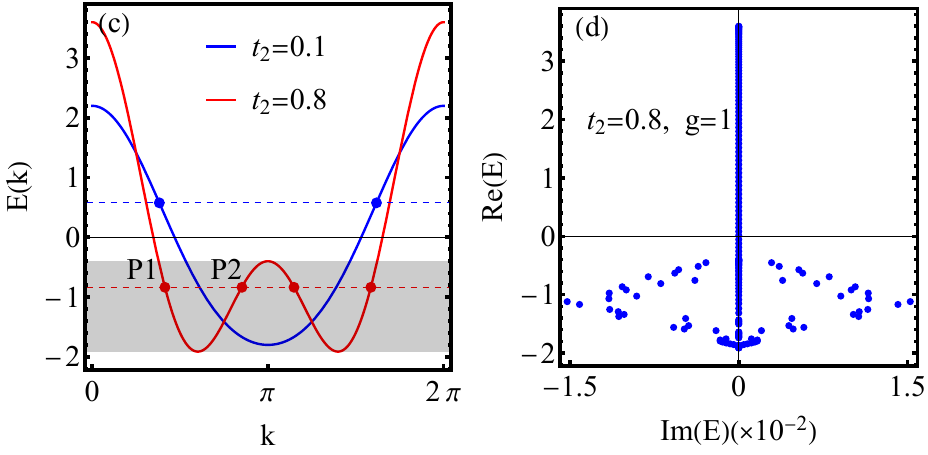}
\end{tabular}
 \caption{(a) Phase diagram for an open chain driven non-Hermitian potential $V_2$, where  $P_{\text{com}}$ is used as a phase indicator, and the size is taken as  $L=100$. (b) The eigenvalue-resolved $\langle \delta x\rangle_n$ (upper,left ticks, black) and  the imaginary part of eigenvalue (lower, right ticks, orange), where band index is arranged in the ascending order of eigenvalue real part. (c) Band structure of $H_0$. When $t_2>1/4$, the curve has two local minima and the shadow region covers the energy range in which PT breaking is possible. (d) Eigenvaues in complex plane (with $L=100$) where the real part of complex eigenvalues falls into the energy range in (c). All plots are made by taking $t_1=1$.}\label{fig:model2}
\end{figure}

\textbf{Effect of the second perturbation}.The boundary perturbation $V_1$ has been discussed in the main text. It leads to a PT breaking with vanished threshold according to an effective theory. Here, we explore the effect of $V_2$, which could give a finite PT breaking threshold. As plotted in Fig.~\ref{fig:model2} (a), the phase diagram shows that the PT breaking is still launched until $t_2=t_1/4$, which agrees well with the criteria stated in the main text. However, in this case, we find that the PT breaking can be triggered only if the non-Hermitian perturbation is strong enough, corresponding a finite threshold.  For completeness, we also further verified the emergent of scale-free localization and the (real part) energy window for PT-broken modes, as shown in Fig.~\ref{fig:model2} (b-d).  Especially, due to that both $H_0$ and $V_1$ respect inversion symmetry, the eigenvalue-resolved mean position $\langle x\rangle_n$ can not distinguish it from an extended wavefunction, since the wavefunction of PT-broken modes is expected to decay from two boundaries into bulk. Instead, the inversion symmetric scale-free localization can be characterized by
\begin{eqnarray}
\langle\delta x\rangle_n=\sum_{j=1}^L|\psi_{n,j}|^2 |j-L/2|/\sum_{j=1}^L|\psi_{n,j}|^2,
\end{eqnarray}
which reflects the asymmetry of half of the wavefunction. One should be aware of that this quantity is not always so sensitive because some combination of plane wave might be also ``inversion half-asymmetric", e.g., $\sin(kx)$ for $k\ll 1$. Nevertheless, in many cases, the quantity is still useful, which can help us clearly see the scale-free localization with inversion symmetry.   As shown in Fig.\ref{fig:model2} (b), the wavefunction profile of PT-broken modes is indeed ``half asymmetric", in contrast to most PT symmetric modes that are uniform even in half space. This plot again confirms that scale-free localization is a concomitant effect of PT breaking.

Now, we use an effective theory based on symmetry analysis to explain the finite PT breaking threshold.  The second perturbation is inversion symmetric  $\mathcal PV_2\mathcal{P}=V_2$, so it can couple OBC modes with same parity. Suppose the involved modes are $|\psi_1\rangle$, $|\psi_2\rangle$, which respect PT symmetry $\mathcal P\mathcal K|\psi_{1/2}\rangle\propto |\psi_{1/2}\rangle$, the elements of effective Hamiltonian can be shown to be real-valued,
\begin{eqnarray}
(H_{2,eff})_{ab}=\langle\psi_a|V_2|\psi_b\rangle= \langle\psi_a|\mathcal P\mathcal K V_2\mathcal P\mathcal K |\psi_b\rangle=\langle\psi_a|^\ast(\mathcal PV_2\mathcal P) |\psi_b\rangle^\ast=(\langle\psi_a|V_2|\psi_b\rangle)^\ast=(H_{2,eff})_{ab}^\ast, \quad(a,b=1,2)\nonumber\\
\end{eqnarray}
where the potential is real $\mathcal K V_2\mathcal K=V_2$ as it respects both inversion and PT symmetry.  As a result, the effective Hamiltonian between two levels, up to a constant, is given by 
\begin{eqnarray}
H_{2,eff}=\Delta_{12}\sigma_z+g(d_x\sigma_x+d_z\sigma_z+id_y\sigma_y)
\end{eqnarray}
where the gap $\Delta_{12}$ is nonzero (no crossing between modes with same parity) and $d_{x,y,z}$ are real parameters describing the interactions induced by non-Hermitian terms. The PT breaking condition $(\Delta_{12}+gd_z)^2+(gd_x)^2-(gd_y)^2<0$ is only satisfied when $g$ exceeds certain threshold since $\Delta_{12}\neq 0$. The threshold will increase if $\Delta_{12}$ become larger, and the gap $\Delta_{12}$ increase with  the growing of $t_2$, as shown  in Fig.~\ref{fig:staticreal} (b). The is consistent with the phase diagram in Fig.~\ref{fig:model2} (a) where the threshold become larger when $t_2$ is increased.

\bibliography{dirac}